\documentclass[11pt,a4paper]{article}  

\newcommand{\COMMENT}[1]{}

\newcommand{\reff}[1]{}           

\newcommand{\spaceline}[1]{\vspace{9pt}}
\newcommand{\spacex}[1]{\vspace{1cm}}
\newcommand{\spacexx}[1]{\vspace{2cm}}

\newcommand{\kl}[1]{\left\langle} 
\newcommand{\kr}[1]{\right\rangle} 

\newcommand{\BEGINFIGURE}{\begin{figure}[H] \center}         

\usepackage{multicol}  

\usepackage[font=small,margin={9mm,9mm}]{caption} 

\usepackage[margin=2.5cm]{geometry}

\usepackage{graphicx}

\usepackage{amssymb}
\usepackage{amsmath}

\usepackage[utf8]{inputenc}  
\usepackage{url}             

\usepackage[T1]{fontenc}

\usepackage{sidecap}   
\usepackage{here}      
\usepackage{setspace}  

\usepackage{changepage}   

\usepackage[colorlinks]{hyperref}

\usepackage{lineno}

\begin{document}
\setlength\parindent{0pt}
\setlength{\parskip}{6pt}       
\setlength{\overfullrule}{10pt}


\begin{center}
\Large
\textbf{Towards New Parity Rules with Asynchronous Updating Inspired by Fatès's Rule}

~~\\
\large
\textit{Rolf Hoffmann}            

\footnotesize
Technical University Darmstadt, Germany

rolf.hoffmann@tu-darmstadt.de
\end{center}
\normalsize 

\COMMENT{
~~\\
~~\\
\normalsize
\textit{Rolf Hoffmann} \footnote{
Technical University Darmstadt, Germany\\
hoffmann@informatik.tu-darmstadt.de}
and 
\textit{NN2}\footnote{
Address2\\
email2} 
}
\COMMENT{
~~\\
\large
\textbf{Rolf Hoffmann}$^1$, 
\textbf{Ulrich Schachner}$^2$  
 and ?
\textbf{Nazim Fates}$^3$              

\footnotesize
$^1 $Technical University Darmstadt, Darmstadt, Germany\\
hoffmann@informatik.tu-darmstadt.de

$^3$ Address2\\
email2  \\  
$^3$ Address3\\  
email3   
\end{center}
\normalsize 
}



\begin{itemize}
\item[] 	
\textbf{Abstract.}
The parity problem in 1D Cellular Automata is a classifier based on a local rule. If the number of ones in a ring of binary cells is odd then the automaton shall converge to
all-one (11...11) else to all-zero (00...00). The objective is to find rules that are related to Fatès' Balanced Rule F but use less active neighborhoods (actions). 
The method is to analyze the effect of each actions and to find a functioning subset of the  12 actions used in rule F. New rules called HA, HB, HC, HF are defined with 10, 6, 4, 8 actions.  
Rule HC and HF need only three inputs whereas all others need four. The classification time of  the rules was determined by averaging over many simulation runs
with random initial configurations. It is almost the same for the rules HA and F, and is longer for the rules HB, HC, HF.
In addition, the rule HD is defined, which updates three cells in parallel and works faster. In future work, the functioning and performance of the proposed rules should be treated in a more formal way, and related rules (also with different update schemes) could be considered. 


\small
\noindent\textbf{Acknowledgement and Remark.} 
I would like to thank my colleagues 
Nazim Fatès, Barabara Wolnik, Anna Nenja, Pedro P.B. de Oliveira, Gerhard Ha\ss{}linger and Ulrich Schachner 
for the ideas presented in their publications about the parity problem,
or for their comments on this version of the paper,
or for personal discussions regarding this topic and its relation to Markov chains.
This research began in December 2022, motivated by the parity rule proposed by Nazim Fatès
in a preliminary version of \cite{2024Fates}.
\normalsize
\end{itemize}

\COMMENT{
\newpage
\tableofcontents
\newpage
}

\COMMENT{
\vspace{3mm}
\hrule
\begin{multicols}{2}
\tableofcontents
\end{multicols}
\newpage
}

\normalsize
\section{Introduction}
\label{sec:Introduction}

The Cellular Automata (CA) \textit{Parity Problem} is a classic benchmark for evaluating the computational power and complexity of different cellular automaton rules. 
Given is a ring of binary valued cells (1D CA under cyclic boundary conditions) where the number of the $n$
cells is odd. 
The CA system evolution shall converge (i) to the configuration \textit{all--one} ($1^n$) if the initial number of ones is \textit{odd},
and (ii) to the configuration \textit{all--zero} ($0^n$) if the initial number of ones is \textit{even}. 
We can view the system as a classifier that classifies the initial configuration either into $1^n$ or $0^n$, depending on whether the parity of the initial configuration is ``odd'' or ``even''.

This work was mainly motivated by the rules proposed by Fates 
\cite{2024Fates}.
The used CA updating scheme is asynchronous, a pair of cells is selected at random
and their state is then updated depending on their neighborhood. 
More good references  about the parity problem and its solution are 
\cite{2019Ruivo,2025Wolnik,2013Betel,1998Sipper}.

The main objective of this paper is to pave the way for new rules solving the parity problem which
(i) have a similar working principle like Fates' Balanced Rule, and 
(ii) use a smaller set of active neighborhoods, or 
(iii) use less than 4 neighbors. 

\newpage
\textbf{Organization of the paper} 
\begin{enumerate}
	\item 	 
	Fatès's rule $F$ is explained and how it can be described by
  a logical function  which triggers the state transitions.
	(Sect. \ref{sec:FRules})
	
  \item
	The effect of all 16 possible active neighborhoods (actions)
  are analyzed and whether they are candidates for new rules.
	(Sect. \ref{Access pattern and actions})	
	
  \item
  New rules \textit{HA, HB, HC, HF} are defined with less then 12 actions
  compared to rule \textit{F}. 
	In addition, the rule \textit{HD} is proposed, which updates three cells in parallel. 
  (Sect. \ref{Rule HA} -- \ref{Rule HF})
			
 \item
  The performance of the rules is evaluated based on simulations.
	(Sect. \ref{Performance Evaluation})
	
	\item
	Appendix 0:
	Transition graphs with a different set of active neighborhoods show how the
	different rules work for $n=5$.
	
  \item
	Appendix 1: Simulation sequences are shown  in order to understand how the final states evolve step by step, starting with special initial configurations (which include the worst time consuming cases).
	We can observe, that the number $b$ of blocks $01^k0$ or $10^k1$ ($k\geq1$) is  decreasing for all rules.
	
	\item
	Appendix 2: Transitions matrices are given for further studies. 
\end{enumerate}

\section{Fatès's balanced rule}
\label{sec:FRules}

Given is a ring of $n$ cells with states $x_i\in\{0,1\}$, $n$ being \textit{odd}. 

~~~
$x= x_0, x_1, ~...~ , x_i, ~...~ x_{n-1}$.

If a considered cell index $i$ is not in the range $[0 ~.. . ~n-1]$ then it is taken $\textit{mod} ~n$.

We declare 

~~~
$L = x_{i-1}$,~
$Z = x_i$,~
$R = x_{i+1}$,~
$Q = x_{i+2}$                          \hfill \textit{-- abbreviations for the cells in the neighborhood}

~~~
\textit{$ZR=(Z,R)$                       \hfill -- the compound state of a cell pair $(i, i+1)$}

~~~
\textit{$LZRQ=(L,Z,R,Q)$                       \hfill-- neighborhood pattern: state of the neighborhood $(i-1, i, i+1, i+2)$}

~~~
$\bar {Z}=1-Z$, and $\bar {R}=1-R$       \hfill \textit{-- the negations/complements}

\textbf{Fatès's Rule F:}

    ~~~$(Z,R)^{t+1} =
    \begin{cases}
     (\bar {Z}, \bar{R})^t \textit{~~if~~} L\neq Z  \textit{~or~}  R\neq Q \hspace{52mm}\textit{-- flip both states} \\     
     (Z,R)^{t}             \textit{~~otherwise}                          \hspace{68mm}\textit{-- keep old states}\\
    \end{cases}$
 
We call this rule \textit{\textbf{F}}.
The compound state \textit{ZR}  
is  flipped between step $t$ and $t+1$  if the activating logical function 

~~~$F= (L\neq Z) \lor (R\neq Q)$ 

is TRUE, otherwise it is kept.
We assume  \textit{fully asynchronous update mode}, meaning that a pair \textit{ZR} is picked at
a random position (uniform distributed) at time $t$ and immediately updated.

This function (Table \ref{TableNazimBalanced}) defines so-called ``active neighborhoods'' 
\cite{2024Fates}
which trigger/enable 
the associated operations (state transitions).
Here, we will call an active neighborhood also  \textit{``action''}. 
We can write $F$ as a logical function,
a disjunction of logical terms (conjunctions over the boolean neighborhood variables
$LZRQ$).
For example, for

~~~``$(\bar{L} \land{Z} \land{R} \land{Q})
\lor
(L \land{\bar{Z}} \land{R} \land{\bar{Q}})$''
we can write ``\texttt{0111 + 1010}''.  

In this alternative notation, 
$F$ can be written as an OR-sum of actions (using the OR operator ``+'' instead of ``$\lor$'', 
which is a common notation for describing combinatorial logic): 

~~~$F(LZRQ) = \texttt{0001+0010+0100+0101+0110+0111+1000+1001+1010+1011+1101+1110}$ 

~~~$= 1 + 2 + 4 + 5 + 6 + 7 + 8 + 9 + 10 + 11 + 13 + 14$  \hfill -- \textit{represented by their decimal equivalents}

or in \textit{minimized} form (through joining neighboring terms in the 4-cube, like 
\texttt{0100+0101 = 010*}) as

~~~$F(LZRQ) =  \verb|01** + 10** + **01 + **10| ~=~$
\Large 
\(	\substack{\texttt{01**}\\
							\texttt{10**}\\
							\texttt{**01}\\
							\texttt{**10}											
						 }
\)
\normalsize

where '\texttt{*}' denotes a Don't Care variable. 
Alternatively we can define the rule by the corresponding table (right side of the last equation) which can emphasize symmetries.

\begin{table}[H]
\center
\caption{(a) This table defines the 16 possible \textit{actions} 
(active neighborhoods) $k$ with the  flip-operation $(Z,R) \rightarrow (\overline{Z},\overline{R})$, and
Rule $F$ uses 12 of them.  
The changes in the number of ones in the pattern is given by $\Delta e$,
and the changes in the number of blocks is given by $\Delta b$ (blocks are explained in Sect. \ref{Access pattern and actions}).
(b) The table of the minimized function $F$ uses the don't care symbol '*',
meaning that the corresponding value can be either 0 or 1.
}

(a)\hfill~~

\scriptsize

\begin{tabular}{c|c|c|c|c|l|r|c|c}
    		
 action   &            &    & oper-      &   state                   &  name of action                                &  LZRQ    &$\Delta e$  &  $\Delta b$       \\								
 index k  & L  ZR  Q   &  F & ation      &   change                  &  (comment)                                     &  parity  &$   $        &         \\
\hline                                                                                                                                      
			 0  & 0  00  0   &  0 &  --         &   $0000 \rightarrow 0110$ &  -- not allowed --                             &even     &$+2 $        &  $ +1  $ \\           
			 1  & 0  00  1   &  1 &  flip       &   $0001 \rightarrow 0111$ &  pl (plus 2, extend 1-block to the left)       &odd      &$+2 $        &  $     $ \\
			 2  & 0  01  0   &  1 &  flip     	&   $0010 \rightarrow 0100$ &  L1 (Left shift of 1 enclosed by 0s)           &odd      &$   $        &  $     $ \\
			 3  & 0  01  1   &  0 &  --       	&   $0011 \rightarrow 0101$ &  Sa (Scatter a)                                &even     &$   $        &  $ +1  $  \\
			 4  & 0  10  0   &  1 &  flip     	&   $0100 \rightarrow 0010$ &  R1 (Right shift of 1 enclosed by  0s)         &odd      &$   $        &  $     $\\
			 5  & 0  10  1   &  1 &  flip     	&   $0101 \rightarrow 0011$ &  JR (Join to the Right)                        &even     &$   $        &  $ -1  $\\
			 6  & 0  11  0   &  1 &  flip     	&   $0110 \rightarrow 0000$ &  M (Minus 2)                                   &even     &$-2 $        &  $ -1  $  \\
			 7  & 0  11  1   &  1 &  flip     	&   $0111 \rightarrow 0001$ &  ml (minus 2, shorten 1-block from the left)   &odd      &$-2 $        &  $     $\\
																																			                                                                                  
			 8  & 1  00  0   &  1 &  flip     	&   $1000 \rightarrow 1110$ & pr (plus 2, extend 1-block to the right)       &odd      &$+2 $        &  $     $ \\
			 9  & 1  00  1   &  1 &  flip     	&   $1001 \rightarrow 1111$ & P (Plus 2)                                     &even     &$+2 $        &  $  -1 $    \\
			10  & 1  01  0   &  1 &  flip     	&   $1010 \rightarrow 1100$ & JL (Join to the Left)                          &even     &$   $        &  $  -1 $ \\
			11  & 1  01  1   &  1 &  flip     	&   $1011 \rightarrow 1101$ & R0 (Right shift of 0 enclosed by 1s)           &odd      &$   $        &  $     $  \\
			12  & 1  10  0   &  0 &  --       	&   $1100 \rightarrow 1010$ & Sb (Scatter b)                                 &even     &$   $        &  $  +1 $  \\
			13  & 1  10  1   &  1 &  flip     	&   $1101 \rightarrow 1011$ & L0 (Left shift of 0 enclosed by  1s)           &odd      &$   $        &  $     $ \\
			14  & 1  11  0   &  1 &  flip     	&   $1110 \rightarrow 1000$ &  mr (minus 2, shorten 1-block from the right)  &odd      &$-2 $        &  $     $ \\
			15  & 1  11  1   &  0 &  --       	&   $1111 \rightarrow 1001$ &  -- not allowed --                             &even     &$-2 $        &  $   +1$  \\
                                                                                                                      
\end{tabular}

\normalsize
\vspace{5mm}
(b)\hfill~~

\noindent
\scriptsize
\begin{tabular}{@{}c|c|c}        
            
             L      ZR     Q  & F  & operation \\
\hline                             
             0      1*     *  & 1  & flip \\
             1      0*     *  & 1  & flip \\
             *      *0     1  & 1  & flip \\
             *      *1     0  & 1  & flip \\
\end{tabular}

\label{TableNazimBalanced}
\end{table}
\normalsize

\section{New rules}


\subsection{Blocks and access pattern}
\label{Access pattern and actions}

\textbf{Blocks.}
A configuration or pattern can be viewed as a cyclic sequence of alternating 1- and 0-blocks.
A 1-block is a sequence of 1s enclosed by 0s ($01^{k\geq1} 0$),
and 
a 0-block is a sequence of 0s enclosed by 1s ($10^{k\geq1} 1$).
Note that the number of 1-blocks and 0-blocks is the same, and the  fixed points $1^n$ and $0^n$ don't form a block
because $1^n / 0^n$ are not enclosed by $0/1$.
The number $b$ of blocks is related to the number of bit-changes which is $2b$.
We aim at rules that conserve the parity and which converge from
a pattern with $b \leq (n-1)/2$ blocks to the fixed points with $b=0$.
\textit{Examples}: $b(11111)=0, b(00100)=1, b(100111011)=2, b(1010101)=3, b(0101010)=3.$

\newpage
\textbf{Access Pattern.} 
An important property of a rule is the number and location
of the cells that provide input values, and
the number and location of the cells whose states are modified. Therefore we 
 relate an \textit{access pattern} $\omega$ (or \textit{read/write pattern}) to 
the rule's  neighborhood $\{-m_1,\ldots,-1,0,1,\ldots,+m_2\}$, 
where $m=1+m_1+m_2\geq1$ is the size of the neighborhood.
We assume that the origin ``0'' of the neighborhood is associated with 
the actual selected index $i$ for processing the rule, which  here is the cell $Z$.
 
We describe the access pattern as a string $\omega = \omega_0, \omega_1, ..., \omega_{m-1}$.
The string symbols are  $\omega_i \in \{r,w,*\}$,
where $r$ means \textit{read-access}, $w$  \textit{write-access}, and *  \textit{no-access}.
Note that not all actions need to make use of the access rights 
assigned to the variables of the whole rule. 
More specifically, if at least one action requires a write-access on a cell $i$ of the neighborhood, 
the whole rule requires it. If at least one action requires a read-access on a cell $j$ of the neighborhood,
and no other action requires a write-access on it, the whole rule requires
a read-access on it. Write-access on a cell allows also a read-access.

The associated access pattern of a rule is $\omega (RULE)$, 
and we may use the notation $RULE^\omega$ to express this feature for a rule. 

\textit{Examples:} Rule $F$ has the access pattern $\omega(F)=rwwr = rw^2r$,
and we may write $F=F^{rw^2r}$.  
The elementary CA rule ECA 30 is $L:= L ~\texttt{xor}~ (Z \lor R)$, it has the access pattern
$rwr$, because the rule reads $L, Z, R$ and writes into the center cell $Z$.

\subsection{The effect of the possible actions depending on the LZRQ neighborhood}
Mainly we are in search for new rules that \textit{work} (correctly converging to $0^n, 1^n$) with less than the 12 actions used by Fatès rule.
The idea is to find working rules checking all subsets of the 
16 input combinations of LZRQ.
First, the effect of the different actions is investigated (see Table 1 and 2).

\begin{table}[H]
\center
\caption{
The set of all possible actions (active neighborhoods).
Actions 1, 9, 8 add a 11-bit-pair  to the actual  selected neighborhood LZRQ, and actions 7, 6, 14 subtract  a 11-bit-pair. 
The actions 10, 5 join two separated 1-bits to a 11-pair.
Action 11/5 shifts Z=0/1 to the right. 
Action 2/13 shifts Z=1/0  to the left.
Action 3 and 12  separate a 11-pair to 101.
See also Table 1.}
\includegraphics[width=0.9\textwidth]{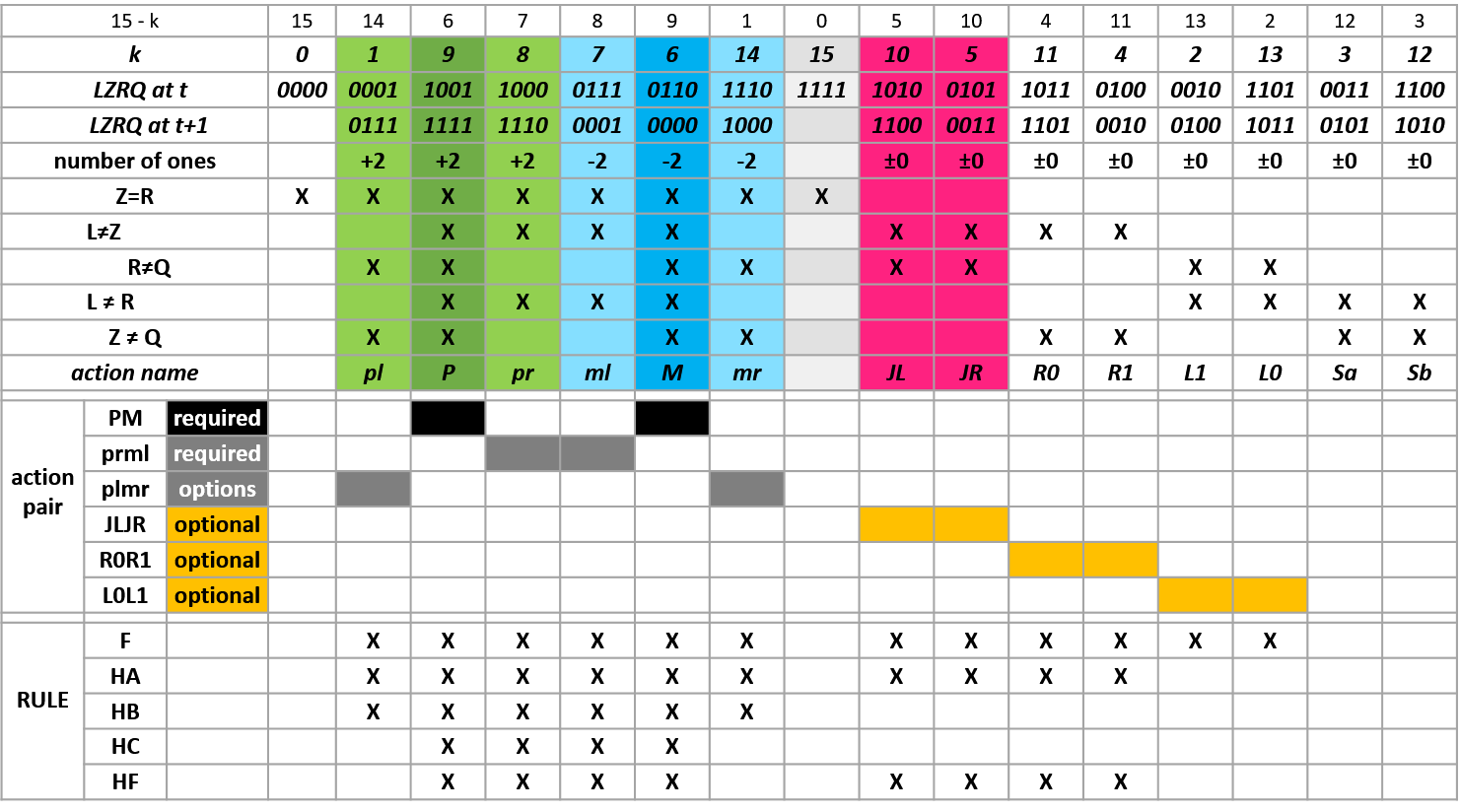}

\label{ActionsNew}
\end{table}   

\begin{itemize}
  \item Actions 0 and 15 are forbidden, because we require stable $0^n$ and $1^n$ configurations, the fixed points for classification, therefore the state pair \textit{ZR} is kept and not flipped.

	\item ``PLUS'' actions \textit{P, pl, pr} flip $ZR = 00 \rightarrow 11$ (adding two 1-bits, subtracting two 0-bits).

			\begin{itemize}
			\item 
      Action \textit{P} ($1001 \rightarrow 1111$)
      eliminates the 0-block $10^21$ and connects the two 1s 
      to one larger 1-block.  
      Thereby the number $b$  of blocks is decreased by 1 ($\Delta b=-1)$
      and the number $e$ of ones increases by 2 ($\Delta e=+2$).

      \item
      Actions \textit{pl} and \textit{pr}     
      conserve the 
      number of blocks ($\Delta b=0)$.
      The number $e$ of ones increases by 2 ($\Delta e=+2$), thereby
      one 1-block is enlarged,

		\end{itemize}
  
  \item ``MINUS'' actions \textit{M, ml, mr} flip $ZR = 11 \rightarrow 00$ (subtracting two 1-bits, adding two 0-bits).

		\begin{itemize}
			\item Action \textit{M} ($0110 \rightarrow 0000$)
      eliminates the 1-block $01^20$ and connects the two 0s 
      to a larger 0-block.  
      Thereby the number of blocks is decreased by 1.
      
      \item
      Action \textit{ml} and \textit{mr}     
      conserve  the number of blocks.
      The size of one 1-block is reduced by 2. 
    
		\end{itemize}
  
  \item ``SWAP'' actions 
\textit{JL, JR,  L0, L1, R0, R1, Sa, Sb}
	swap the bits of the bit pair $ZR = 01 \leftrightarrow 10$,
	and they are number conserving ($\Delta e=0$).

  \begin{itemize}
    \item 
    ``JOIN'' actions \textit{JL, JR} are very useful, because separated bits of the same value are joined
		and thereby the number of blocks is reduced ($\Delta b=-1$). 
    
    $0101 \rightarrow 0011$,   $1010 \rightarrow 1100$. 
    
    \item
    ``SHIFT'' actions 
		\textit{L0, L1, R0, R1}
		shift a 0 or a 1 in the middle to the left or right. ($ZR = 01 \leftrightarrow 10$).
		Although 
		$\Delta e=0$ and $\Delta b=0$, these actions introduce some sort of ``noise'' or ``random walk''
		which may improve the classification time.
		The 4-bit neighborhood patterns are mirrored, e.g. $1011 \leftrightarrow 1101$. 		
		
    \item 
    ``SCATTER'' actions 
		\textit{Sa, Sb}
		should not be used because bit pairs of the same value are
    separated:     
    $0011 \rightarrow 0101$, $1100 \rightarrow 1010$. These actions reverse the useful effects of the JOIN actions.
        (Some tests have shown that the classification time is much longer if these action  are included in a rule, 
    but still the classification can be performed with them.)

  \end{itemize}
  
\end{itemize}

Now the questions is to find other rules with less than 12 actions. 
For a \textit{balanced} rule which treat the odd and even classifications in a fair complementary way, 
for each action $LZRQ$ the  bitwise negated action
$\overline{L}\overline{Z}\overline{R}\overline{Q}$ has to be included. 
In other words, a balanced rule contains for each action $k$ also action $15-k$, the one's complement to 
$15 = 1111_2$

A rule can be \textit{``neighborhood--symmetric''} with respect to an equal treatment of the left/right neighbors. 
We define a rule as neighborhood--symmetric 
if the rule contains only symmetric action pairs ($LZRQ=abcd$, $LZRQ=dcba$),
 or self-symmetric actions $LZRQ=abba$
where $(a,b,c,d) \in\{0,1,\texttt{*}\}$.
For example, symmetric action pairs are: $(0001,1000), (0111, 1000), ...$,
and self-symmetric actions are 1001, 0110.
Rule $F$ is neighborhood-symmetric. 

\subsection{Candidate rules}

During the investigations it turned out (by studying transition graphs the onbes shown in Appendix 0, and suggested by simulation results)
that working rules need the actions 
\textit{PM} and either \textit{plmr} or \textit{prml} or both.
In addition, the actions 
\textit{JLJR}, \textit{R0R1}, or \textit{L0L1} can be applied in any combination, where 
(see Tables \ref{ActionsNew}, \ref{RuleSet})

\(
~~~\textit{PM = P+M, plmr = pl+mr, prml = pr+ml} 
\)

\(
~~~\textit{JLJR = JL+JR, ~R0R1 = R0+R1, ~L0L1 = L0+L1} .
\)

The collection of the 24 suggested candidate rules (Table \ref{ActionsNew}) can formally be described as

\textit{~~~%
\(
\mathcal{R} = \{R|R\subseteq \{PM,plmr,prml,JLJR,R0R1,L0L1\}\land PM\in R\land (plmr\in R\lor prml\in R)\}
\)
}

or

\textit{~~~$\mathcal{R} = \left\{ \{PM\} \cup Y \cup Z \;\middle\vert\; (Y, Z) \in \mathcal{P}^+(\{plmr, prml\}) \times \mathcal{P}(\{JLJR, R0R1, L0L1\}) \right\}$}.

These formulas mean: Every rule (defined as a set of actions) in the collection is built by choosing: The element \textit{PM} (mandatory). 
A choice from the three combinations of \textit{plmr} and \textit{prml}.
Any choice from the 8 subsets of \(\{\textit{JLJR, R0R1, L0L1\}}\) 
(including the empty set $\emptyset$). 

Most of these rules were simulated and showed the required classification up to a ring size of $n=39$.

The SCATTER actions can also be applied, but they slow down the classification time significantly.

\begin{table}[H]
\center
\caption{
The candidate rules. Most of them were simulated and showed the required behavior.}
\includegraphics[width=0.9\textwidth]{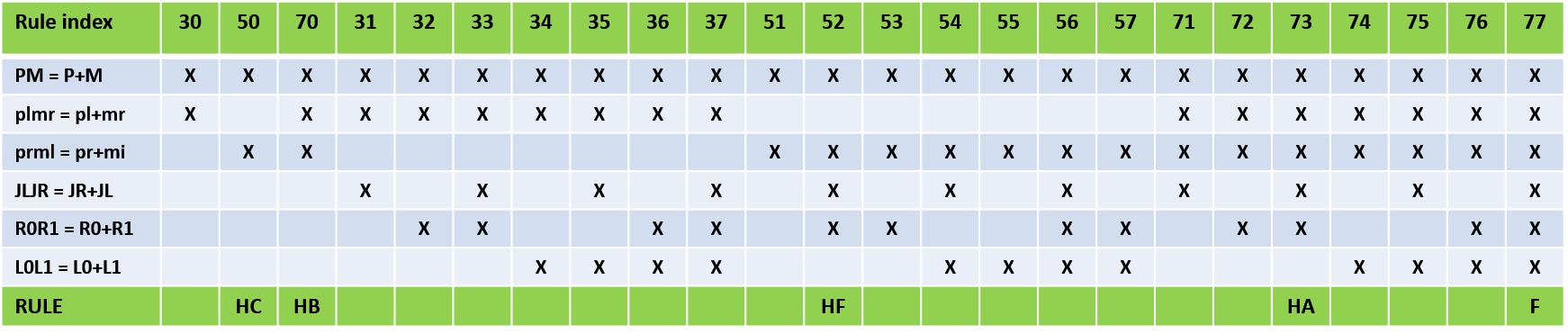}
\label{RuleSet}
\end{table}   

\subsection{Rule HA}
\label{Rule HA}

The new rule \textit{HA}$^{rw^2r}$ uses 10 actions. It is not neighborhood-symmetric. 
Compared to  rule \textit{F}, it does not use the actions \textit{L0L1}, and
we can express this by the notation \textit{$\textit{HA} = F - L0L1$.}

~~~$\textit{HA} =  \textit{PM} + \textit{pmlr} + \textit{prml} + \textit{JLJR} + \textit{R0R1}~=~$
\Large 
\(	\substack{\texttt{01**}\\
							\texttt{10**}\\
							\texttt{*001}\\
							\texttt{*110}											
						 }
\)
\normalsize

\normalsize
~~~$\textit{HA} = (L \neq Z) \lor [(Z=R) \land (R \neq Q)]$

\subsection{Rule HB}
\label{Rule HB}

The new rule $\textit{HB}^{rw^2r}$ uses 6 actions, it is neighborhood-symmetric.

~~~$\textit{HB} = \textit{PM} + \textit{plmr} + \textit{prml}~=~$ 
\Large 
\(	\substack{\texttt{*001} ~\rightarrow~ \texttt{*111}\\
							\texttt{*110} ~\rightarrow~ \texttt{*000}\\
							\texttt{100*} ~\rightarrow~ \texttt{111*}\\
							\texttt{011*} ~\rightarrow~ \texttt{000*}											
						 }
\)
\normalsize

~~~$\textit{HB} =(Z=R) \land [(L\neq Z) \lor (R \neq Q)])$

The JOIN and SHIFT actions are not used.

\subsection{Rule HC}
\label{Rule HC}

The rule with the minimal number of 4 actions is

$~~~\textit{HC} = \textit{PM} + \textit{prml}~=~$
\Large
\(
              \substack{ \texttt{100*}  ~\rightarrow    \texttt{~111*} \\ 
							           \texttt{011*}  ~\rightarrow    \texttt{~000*}
												}												
\)
\normalsize

$~~~HC = (L \neq Z) \land (Z = R)$ 

This rule is \textit{not} neighborhood--symmetric. 
The number of rule inputs is only 3, because the state $Q$ is not taken into account.
Therefore the access-pattern is $\omega(HC) = rww*$.
The JOIN and SHIFT actions are not used. 

It is not obvious how the rules \textit{HB} and \textit{HC} can work without 
JOIN and SHIFT.
We observe
(see simulations Sects. \ref{SimRuleHB} and \ref{SimRuleHC})
that, even for the worst initial case, the evolution can successfully proceed,
starting with a unique state transition 
   $00 \rightarrow 11$ (Table \ref{simruleHB} c,d; $u=1$, and Table  \ref{simruleHC} c,d), 
or $11 \rightarrow 00$ (Table \ref{simruleHB} b; $u=1$,   and Table  \ref{simruleHC} b). See also the transition graphs in Appendix 0.
Thus the information flow  required for convergence is ensured  indirectly via the actions 
$\textit{ZR} = 00 \leftrightarrow 11$.
 
We can shorten the rule because only the inputs \textit{LZR} are used with the access pattern $rww$:

~~~ 
\(\textit{HC} ~=~\)
\Large
\(
              \substack{ \texttt{LZR}           \texttt{~~~~~~~}\\
						           	 \texttt{100}  ~\rightarrow \texttt{~111}\\
                         \texttt{011}  ~\rightarrow \texttt{~000}
											 	}											
\)

\normalsize

To explain the following rule variants, we write rule $HC$, supplemented by the column identifiers \textit{“abc”} and the access pattern $rww$.
	
~~~	
\(\textit{HC} ~=~\)
\Large
\(
              \substack{ \texttt{abc}           \texttt{~~~~~~~}\\
							           \texttt{rww}           \texttt{~~~~~~~}\\
						           	 \texttt{100}  ~\rightarrow \texttt{~111}\\
                         \texttt{011}  ~\rightarrow \texttt{~000}
											 	}											
\)
\normalsize

\subsubsection{HC1: Variant 1 of Rule HC}		

		By interchanging the columns \textit{a} and \textit{c} including their access rights
		we yield the mirrored rule 
				which behaves equivalently if the configuration is mirrored by 
		interchanging  $x_{n-1-i}$ with $x_i$.

~~~
\textit{HC1 = HC$^{mirror} = $}
\textit{HC$^{a\leftrightarrow c}~=~$}		
\Large
\(
              \substack{ \texttt{cba}           \texttt{~~~~~~~}\\
							           \texttt{wwr}           \texttt{~~~~~~~}\\
						           	 \texttt{001}  ~\rightarrow \texttt{~111}\\
                         \texttt{110}  ~\rightarrow \texttt{~000}
											 	}											
\)
\normalsize

\subsubsection{HC2: Variant 2 of Rule HC}	
		By interchanging the columns $a$ and $b$ including their access rights
		we yield the rule 

~~~
$\textit{HC2} = \textit{HC}^{a\leftrightarrow b}~=~$			
\Large
\(
              \substack{ \texttt{bac}           \texttt{~~~~~~~}\\
							           \texttt{wrw}           \texttt{~~~~~~~}\\
						           	 \texttt{010}  ~\rightarrow \texttt{~111}\\
                         \texttt{101}  ~\rightarrow \texttt{~000}
											 	}											
\)
\normalsize

This rule is quite different, because the access pattern is \textit{wrw},
meaning that the cells being updated are separated by a read-only cell. 
Can it work? Unfortunately not, because blocks of zeroes or ones 
($0^k$, $1^k$ with $k\geq2$) remain stable. 
Nevertheless these state transitions can be utilized for the construction of rule \textit{HD}
(Sect. \ref{Rule HD}).

\subsection{Rule HF with three inputs}
\label{Rule HF}

This new rule \textit{HF} rule needs only three inputs like rule \textit{HC}.
It was found by taking only the left or right part of rule \textit{F},
so there are two variants. 
We assume that \textit{ZR} is flipped when the rule becomes true. 

~~~$F= (L\neq Z) \lor (R\neq Q)$  \hfill~~\textit{Rule F}

~~~$\textit{HF1} = (L\neq Z)$  \hfill~~\textit{Variant 1: left part of Rule F}

~~~$\textit{HF2} = (R\neq Q)$  \hfill~~\textit{Variant 2: right part of Rule F}

Rule \textit{HF1} triggers the state transitions: 

~~~\(\{100 \rightarrow 111, 
101 \rightarrow 110, 
010 \rightarrow 001, 
011 \rightarrow 000\}\) .

Written in  the former used notation, we yield

~~~\(\textit{HF1} ~=~\)
\Large
\(
              \substack{ \texttt{LZR}           \texttt{~~~~~~~}\\
							           \texttt{rww}           \texttt{~~~~~~~}\\
						           	 \texttt{10*}  ~\rightarrow ~\texttt{1}\bar{Z}\bar{R} \\
                         \texttt{01*}  ~\rightarrow ~\texttt{0}\bar{Z}\bar{R}
											 	}																						
\)
\normalsize	
\(~=~\)
\Large
\(
              \substack{ \texttt{LZR}           \texttt{~~~~~~~}\\
							           \texttt{rww}           \texttt{~~~~~~~}\\
						           	 \texttt{100}  ~\rightarrow ~\texttt{111} \\
						           	 \texttt{101}  ~\rightarrow ~\texttt{110} \\
                         \texttt{010}  ~\rightarrow ~\texttt{001} \\												
                         \texttt{011}  ~\rightarrow ~\texttt{000}
											 	}																						
\)
\normalsize .

\subsection{Rule HD with three write-cells}
\label{Rule HD}
		
Rule \textit{HD} belongs to another category of rules in which three cells are updated in parallel,
the access pattern is $w^3=www$.
It was found by ``adding'' the rules $\textit{HC} + \textit{HC1} + \textit{HC2} = \textit{HD}$. Then it was realized that it could 
also easily be designed, because 
(i) 000 and 111, which are part of the absorbing states  $0^n$ and  $1^n$, are not changed, 
(ii) the \textit{odd} states with one 1-bit (001, 010, 101) are changed into 111,
driving the evolution to the global state $1^n$, 
and
(iii) the \textit{even} states with two 1-bits (or one 0-bit: 110, 101, 011) are changed into 000,  
driving the evolution to the global state $0^n$.

$\textit{~~~HD}~=~$
\Large
\(
              \substack{ \texttt{001}  ~\rightarrow \texttt{~111}\\
							           \texttt{010}  ~\rightarrow \texttt{~111}\\
							           \texttt{100}  ~\rightarrow \texttt{~111}\\
						           	 \texttt{110}  ~\rightarrow \texttt{~000}\\
												 \texttt{101}  ~\rightarrow \texttt{~000}\\
                         \texttt{011}  ~\rightarrow \texttt{~000}
											 	}											
\)
\normalsize

Other rules with the  same access pattern $w^3$ but with less actions (and a lower time-efficiency) are
\textit{HC} + \textit{HC1}, \textit{HC} + \textit{HC2}, and \textit{HC1} + \textit{HC2}.

Following the working principle of rule \textit{HD}, similar rules with $k$ neighbors and
the access pattern $w^k$ ($k=5, 7, ... $)
could be defined, which map every \textit{k}-bit string with odd parity to $1^k$,
and to $0^k$ for even parity.

\section{Performance evaluation}
\label{Performance Evaluation}

The used performance measure is the \textit{average classification time} $T_{avr}$ ,
averaged over the mean classification times $T(InitialConfig)$ 
for all uniformly distributed random initial configurations.
It was obtained by performing $N_{run}$ simulation runs and counting the number of
micro-time steps until classification,
starting from a random initial configuration, or with a special initial configuration 
when only the behavior for such a configuration was of interest.
For $n= 5, \ldots ,27$ the number of runs was  $N_{run}=50,000$,  20,000 for $n=29,\ldots,39$,
and 10,000 for $n > 39$.

Here, we measure time in \textit{micro-time} steps. Alternatively, a time step can be defined as consisting of a sequence of $n/W$ micro-time steps, where $W$ is the number of cells being updated in parallel. This definition is useful when thinking in terms of generations---as is the case with synchronous updating.

\subsection{Average time for the rules F, HA, HB, HC}

We know that the time complexity of Fatès' rule $F$ is $T_{avr} = O(n^3)$.
Therefore we can divide  $T_{avr}$ by $n^3$ in order to get a graph
that converges to a horizontal line representing a constant (Fig. \ref{TimeRuleFHAHBHC}).
Rule $\textit{HA}$ shows the same time complexity as Fates' rule $F$ with
very small differences. 
Until now it is not clear whether $T_{avr}$ of \textit{F} and \textit{HA} are exactly the same.
Rule $\textit{HB}$ is slower but it is almost constant even for small $n$.
Rule $\textit{HC}$ is the slowest but still quite efficient. 
As the graphs for $HB$ and $HC$ seem to converge to a constant
we can assume  the time complexity $T_{avr}=O(n^3)$ also, but a proof is needed.
Further research is necessary to prove that the new rules work for any odd ring size, and to 
compute $T_{avr}$ exactly. 

\BEGINFIGURE
\center

\includegraphics[width=0.6\textwidth]{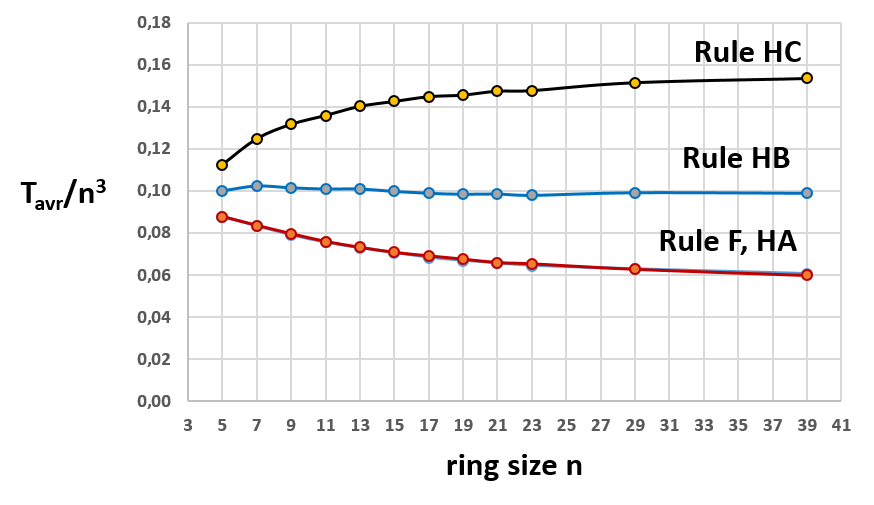}

\caption{
The average classification time $T_{avr}$  divided by $n^3$.
Rule HA shows same time (within a very small range) as Fates' rule F.
Rules HB and HC are slower but seem 
to have the same time complexity of the order $O(n^3)$.
}

\label{TimeRuleFHAHBHC}
\end{figure}   

\subsection{Maximal  vs average mean classification time for the rules F and HA}

Simulation results and some preliminary studies with Markov chains 
suggest that for rules \textit{F} and \textit{HA} the worst case appears for the initial configurations
 (or their cyclic shift equivalents):

~~~$10^{n-1}$ if the parity is odd, and $01^{n-1}$ if it is even,
as depicted  in Fig. \ref{TimeMaxRuleFHA}.

\BEGINFIGURE
\center

\includegraphics[width=0.6\textwidth]{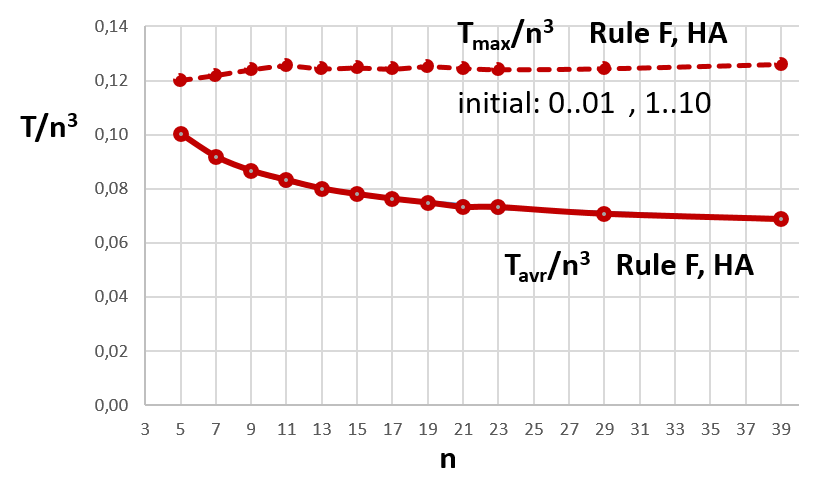}

\caption{
The maximum of the mean classification times (dotted line, for the ``worst'' initial configurations) compared to the average of the mean times (for random initial configurations). 
It is the same for the rules F and HA (with very small differences in the simulation results).
The maximum time is obtained if the initial configuration is $0^{n-1}1$ or $1^{n-1}0$ (or their cyclic shifted equivalents). 
}

\label{TimeMaxRuleFHA}
\end{figure}  

\subsection{Maximal vs average mean classification time for the rules HB and HC}

First, it was assumed that  the worst (most time consuming) initial configurations for rules \textit{HB} and \textit{HC} 
are the same than the ones for  \textit{F} and \textit{HA} ($10^{n-1}$ and $01^{n-1}$).
A test revealed that this assumption is incorrect; in fact, these cases perform better than average rather than worse 
(Fig. \ref{TimeHBHC}a).

\BEGINFIGURE
\center
(a)
\includegraphics[width=0.43\textwidth]{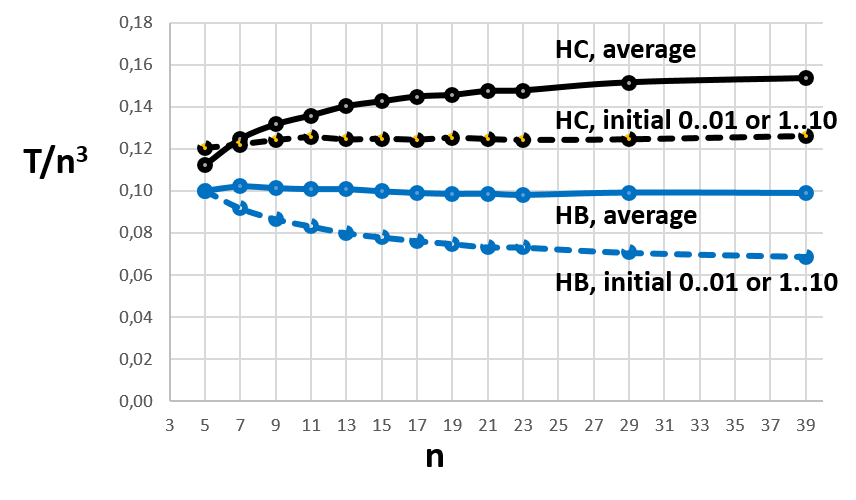}
(b)
\includegraphics[width=0.40\textwidth]{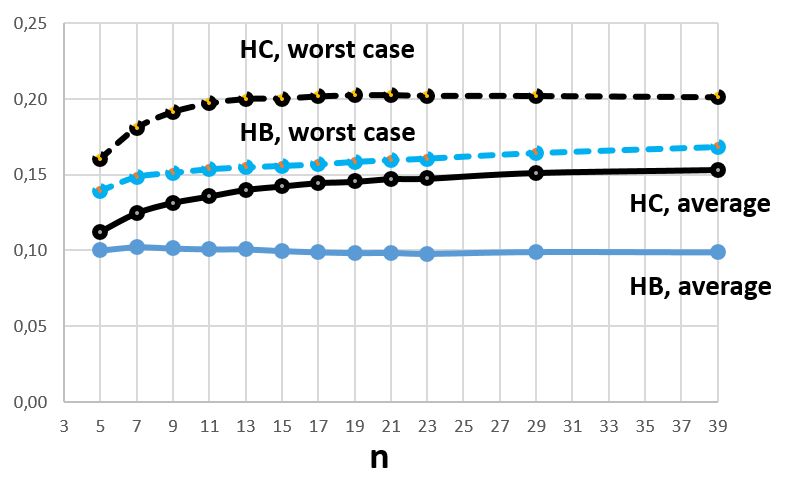}

\caption{
Rules HB vs HC.
(a) The times for the initial configurations $0^{n-1}1$ or $1^{n-1}0$  
(dotted lines) are smaller than the average time.
(b)
Worst time vs average time.
The worst-time initial conditions are 1(10)* for $n= 5, 9, 13, ...$
and 0(01)* for $n=7, 11, 15, ...$ , for the partial problem when the number of of ones is odd. 
}
\label{TimeHBHC}
\end{figure}  

Simulation results and preliminary studies with Markov chains 
(see also Appendix \ref{APPENDIX 0}, Fig. \ref{Graphs}d,e)
suggest that for rules \textit{HB} and \textit{HC} the worst case appears for the initial configurations
 (or their cyclic equivalents):

~~~$1(10)^* \equiv 1(10)^{(n-1)/2}$  for $n = 5, 9, 13, ...$ 
~~and
~~~$0(01)^* \equiv 0(01)^{(n-1)/2}$ for $n = 7, 11, 15, ...$

if odd parity is classified. 
These initial configurations contain a maximal number of blocks.

For the even parity classification, similarly, the negated configurations have to be taken into account. 
The worst-time and the average time are depicted in Fig. \ref{TimeHBHC} b.

In the following section, all possible initial configurations were simulated for $n=7, 9$
in order to confirm these worst-time initial configurations.

\subsection{Time for rule HF} 

Rules \textit{HF1/2} require $T_{avrg}=6184$ micro-time steps for $N=39$,
averaged over 20,000 runs with random initial configurations.
We have $T_{avrg}/n^3 = 0.10425$; this is approximately twice the time (0.059) compared to
Rule \textit{F}.
This result can be roughly explained by the fact that rules \textit{HF1/2} use only half the conditions of Rule \textit{F}.

\subsection{Time for rule HD} 
Rule $HD$ is around $0.07/0.05 = 1.4$ times faster than rule $F$ 
(Fig. \ref{TimeRuleD} vs Fig. \ref{TimeMaxRuleFHA}).
Interestingly, the average mean classification time closely corresponds to the time determined for the initial configuration with the maximum number of blocks.

\BEGINFIGURE
\center

\includegraphics[width=0.6\textwidth]{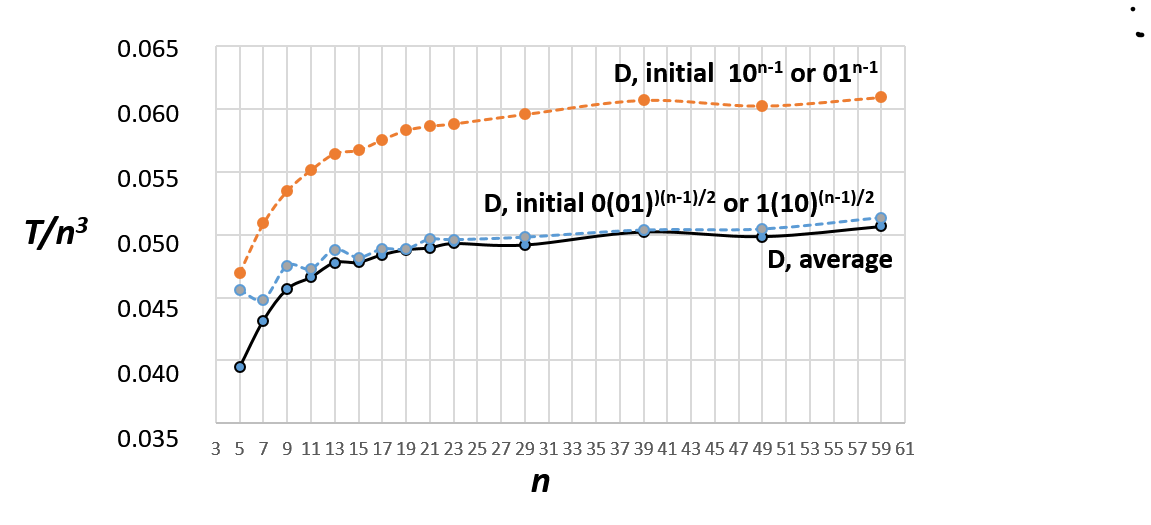}

\caption{
Rule HD. The average mean classification time, and for comparison the times
for the initial configurations (or their cyclic equivalents)  
$10^{n-1}$ or $01^{n-1}$, and
$0(01)^{(n-1)/2}$ or $1(10)^{(n-1)/2}$. 
}

\label{TimeRuleD}
\end{figure}  

\subsection{Rules F, HA, HB, HC: Time for all initial configurations for $n = 7, 9$}

All initial configurations (representatives under cyclic shift) for $n$ odd were considered,
excluding the accepting state $1^n$.
For each configuration 50,000 runs/simulations were performed and the average classification time was measured
(Table \ref{TimesInitsN7} for $n=7$ and Table \ref{SortOrigin} for $n=9$).
The results confirm the worst-time initial configurations,
 $10^{n-1}$/$01^{n-1}$ for the rules \textit{F} and \textit{HA}, and
1(10)*/0(01)* for the rules \textit{HB} and \textit{HC}.

\begin{itemize}
	\item $n=7$ (Table \ref{TimesInitsN7})
  
  There are only 9 representatives (under cyclic shift symmetry).
  State 11 and 13 are mirror-symmetric.  
  The best case is $x^{(t=0)}=x^0 = 31$ for all rules.
  The worst case is $x^0 = 1$ for rule \textit{F} and \textit{HA}, and 21 for the rules \textit{HB} and \textit{HC}.
  The rules \textit{F} an \textit{HB} are neighborhood--symmetric, therefore the 
  mirror configurations 11/13 lead to the same time,
  and it is not necessary to simulate both of them. 
  
 \item $n=9$ (Table \ref{SortOrigin}) 
  
  There are only 29 representatives.
  Seven states are mirror-symmetric: 13/11, 25/19, 41/37, 59/55, 61/47, 109/91, 117/87.
  If a neighborhood--symmetric rule (\textit{F, HB}) is applied, then the mirror-symmetric states yield the same time (with small probabilistic simulation errors).
  The best case is $x^0 = 127$ for all rules.
  The worst case is $x^0 = 1$ for rule \textit{F} and \textit{HA}, and 171 for the rules \textit{HB} and \textit{HC}.
\end{itemize}

\begin{table}[H]
\center
\caption{
Ring size is $n=7$. The classification time (micro-time steps)  $T$ was obtained by averaging over 50,000 simulation runs for each initial configuration.
Instead of simulating all possible configurations, only the
nine representatives were used.
}
\includegraphics[width=0.6\textwidth]{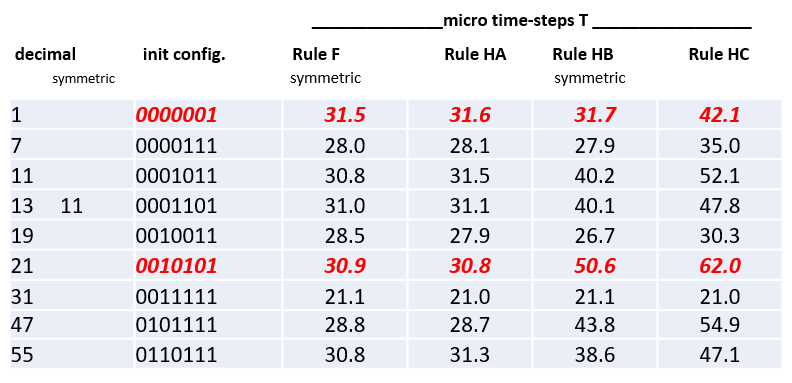}
\label{TimesInitsN7}
\end{table} 

\begin{table}[H]
\center
\caption{
Ring size is $n=9$. The classification time  $T$ was obtained by averaging over 50,000 simulation runs for each
initial configuration. Instead of simulating all possible configurations, only the
29 representatives were used.
}
\includegraphics[width=0.6\textwidth]{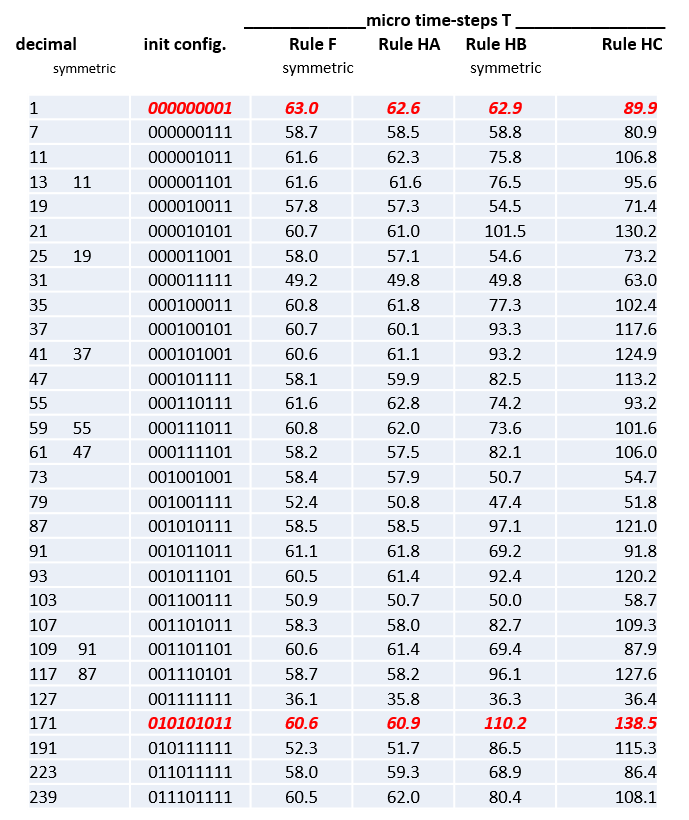}
\label{SortOrigin}
\end{table} 

\section{Conclusion}


The effect of all possible actions (active neighborhoods) was analyzed.
Fatès's balanced rule \textit{F} employs 12 actions.
The newly defined rules \textit{HA, HB, HC,} and \textit{HF} utilize only 10, 6, 4, and 8 actions, respectively.
Rule \textit{HC} requires only three rule inputs,
whereas all other rules use four.
Rules \textit{F} and \textit{HB} are neighborhood-symmetric with respect to their right and left neighbors.
The average mean classification time $T_{avr}$ (expressed in micro-time steps or rule applications) of the new rules was determined via simulation,
yielding $T_{avr} = O(n^3)$ and matching the complexity for rule \textit{F}.
The worst-case initial configurations
are $10^n$/$01^n$ for rules \textit{F}, \textit{HA}, and \textit{HF}
and $0(01)^{(n-1)/2}/1(10)^{(n-1)/2}$ for rules \textit{HB} and \textit{HC}.
The fastest rules are \textit{F} and \textit{HA}, and the slowest rule is \textit{HC}.

In addition, rule  \textit{HD}  was proposed which updates three cells in parallel,
thereby outperforming the rules \textit{F} and \textit{HA}.
To facilitate further analysis using Markov chains, transition matrices were calculated
for rings of sizes 5, 7, and 9, covering all possible initial configurations
that are equivalent under cyclic shift.
Future work should investigate the properties of the proposed rules in greater detail
and consider additional rules with larger neighborhoods or alternative update schemes.



\newpage
\section{APPENDIX 0: Transition Graphs for $n=5$}
\label{APPENDIX 0}

\BEGINFIGURE
\center

\includegraphics[width=0.63\textwidth]{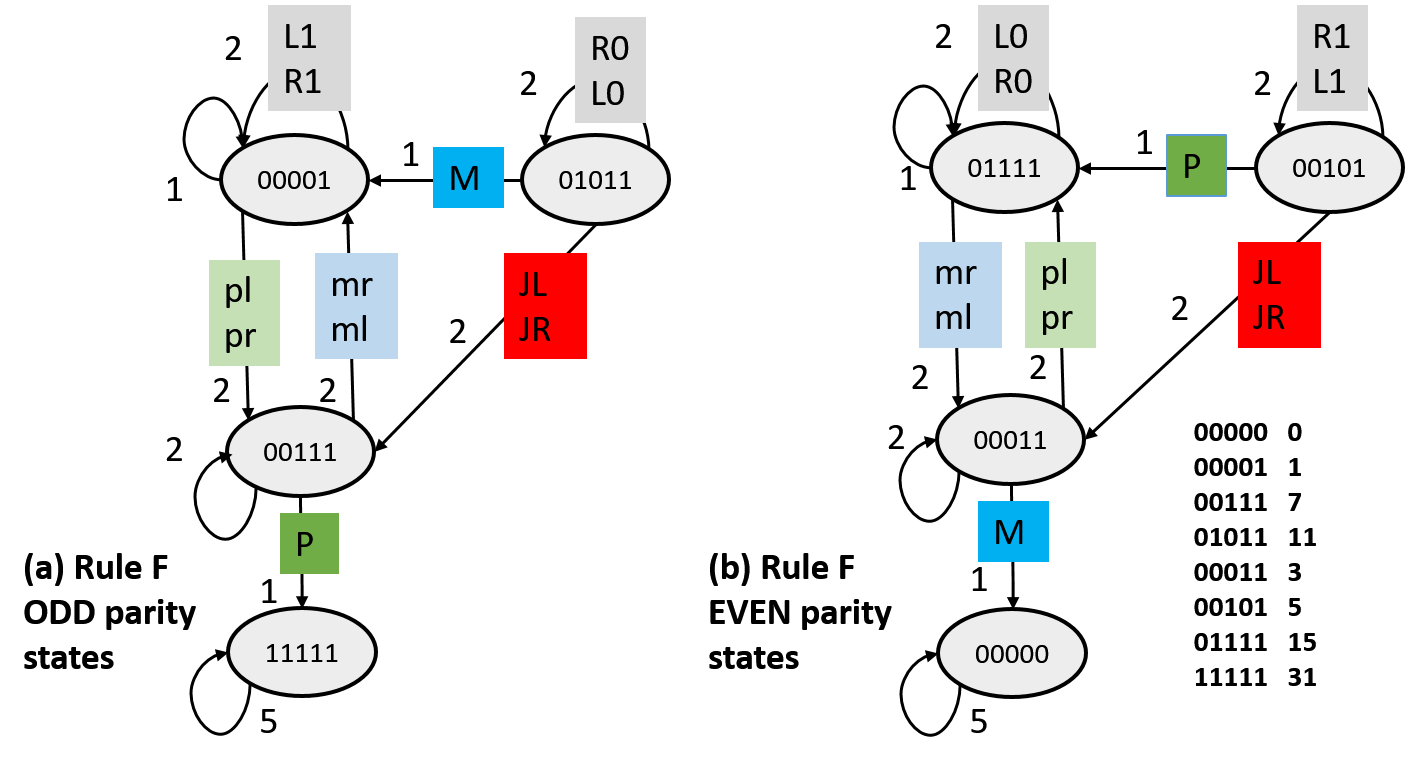}
\includegraphics[width=0.63\textwidth]{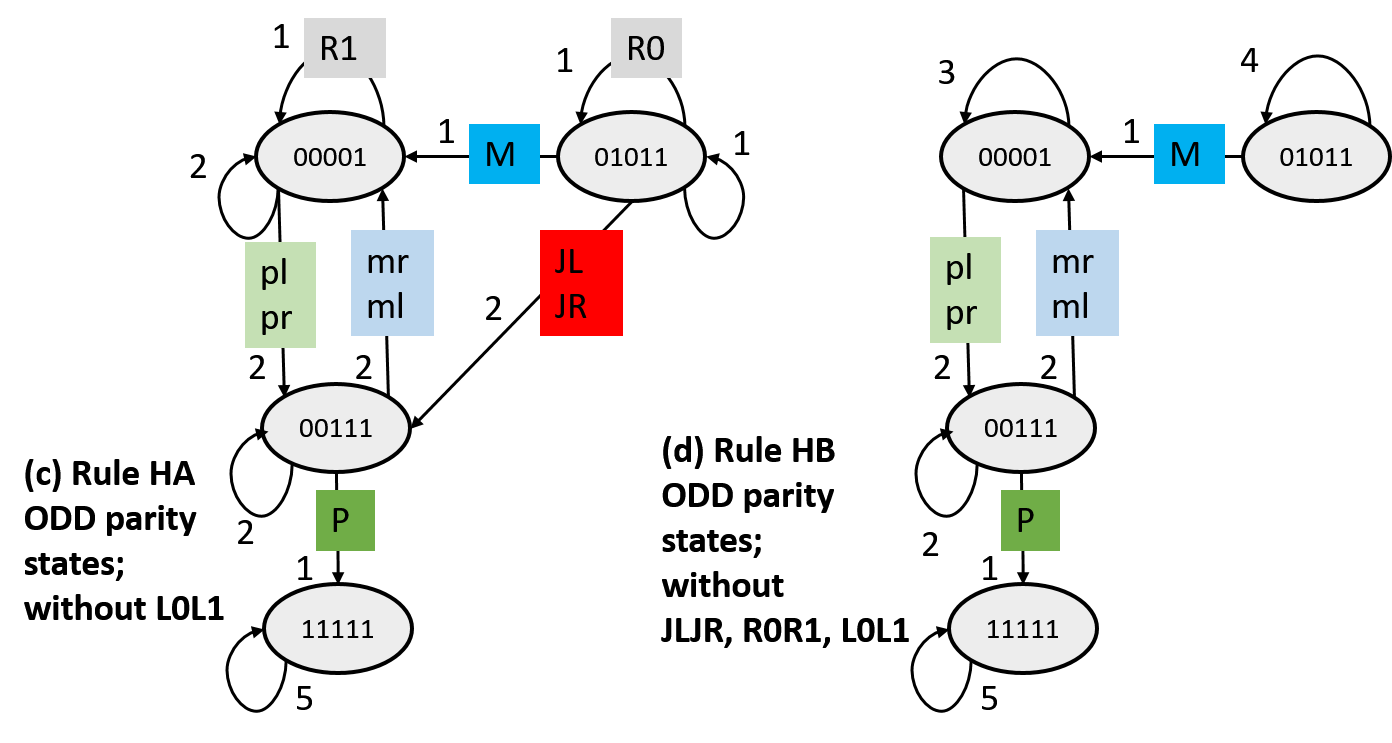}
\includegraphics[width=0.63\textwidth]{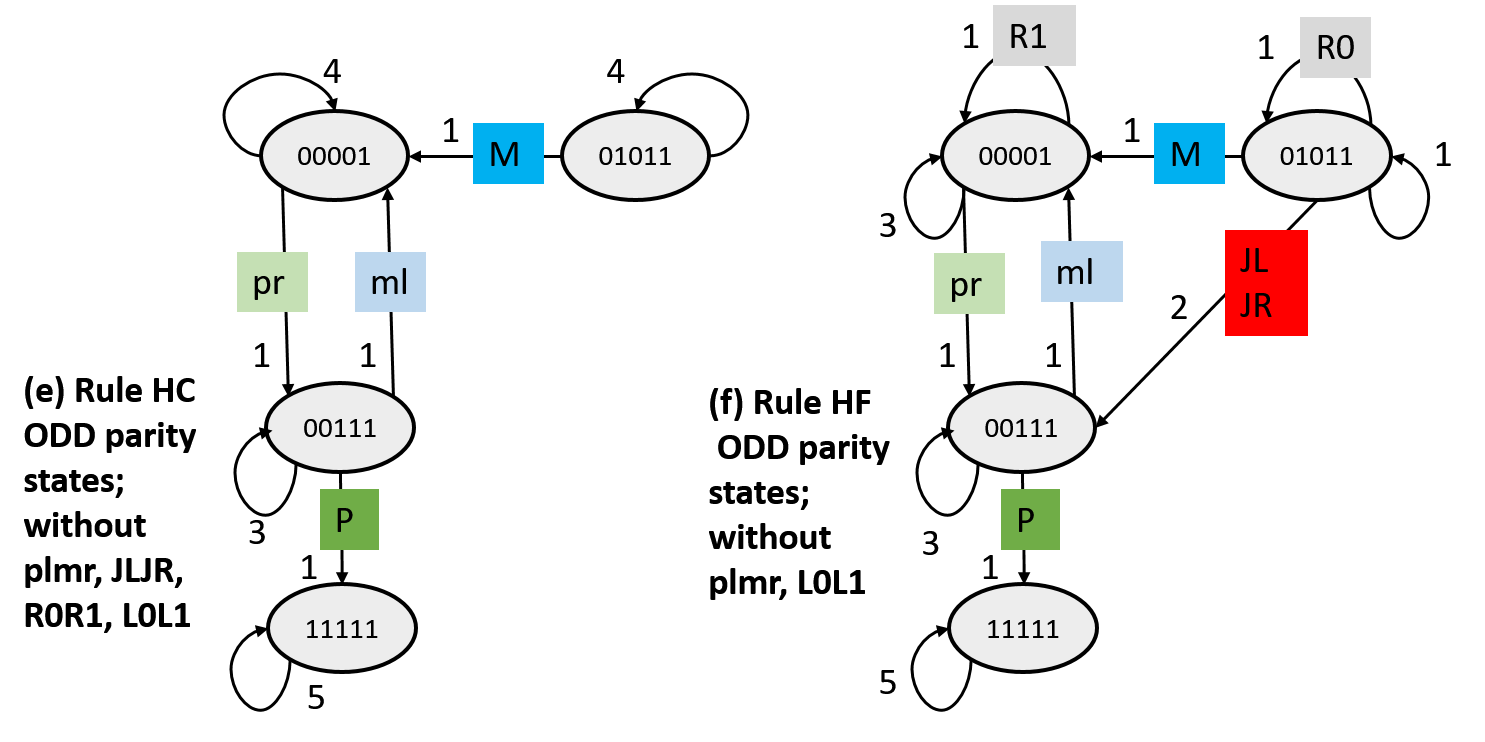}
\includegraphics[width=0.63\textwidth]{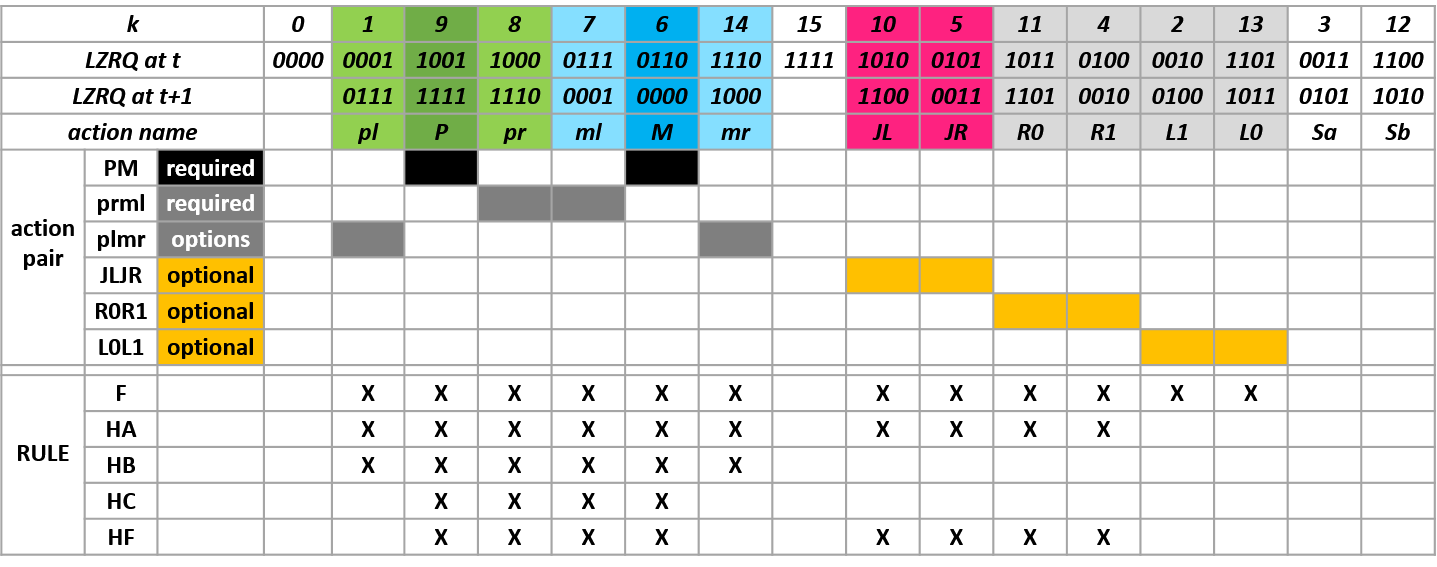}

\caption{
Transition graphs for the rules F, HA, HB, HC, and HF, for $n=5$.
States which are equivalent under cyclic shift are shown only once (represented by their
binary number which has the minimal value).
The numbers at the transitions have to be divided by $n$ giving the transition probability.
Rule F uses all actions (except the scatter actions Sa, Sb). The other rules
are working with less actions, therefore less transitions are triggered. 
Rules HB and HC don't use JLJR, L0L1, and R0R1.
}
\label{Graphs}
\end{figure} 


\newpage
\section{APPENDIX 1: Simulations (Time-Evolutions)}

We show simulations sequences 
for a better understanding how the (odd) classification works step by step for the different rules.
Most interesting are the initial configurations with one block 
($0^{(n-1)/2}~1~0^{(n-1)/2} \equiv 0^{n-1}1$)
and with $(n-1)/2$ blocks ($0(01)^* \equiv 0(01)^{(n-1)/2}  ~\text{or}~ 1(10)^* \equiv 1(10)^{(n-1)/2}$),
because they may need a maximal number of updates
((shortest) path length from the initial configuration to $1^n$ is maximal). 
Shortest sequences (in terms of updates $u$) are shown which appeared during 5000 simulation runs.
For all rules,
the number of ones $e$ is increasing with fluctuations towards $e=n$,
and the number of blocks is decreasing towards $b=0$.

\subsection{Rule F}
\label{SimRuleF}
\begin{table}[H]
\center
\caption{
Rule F. Examples for initial configurations  (i) $0^{n-1}1$ containing one \textit{1-block},   
or (ii) 0(01)* or 1(10)* containing the \textit{maximal number} of 1-blocks.
Only changed configurations are displayed, and not-changed 0 is displayed as ``.'',
and not-changed 1  as ``|''.
\textit{t}: micro-time step; \textit{u}: number of updates; \textit{e}: number of ones; 
\textit{b}: number of 1-blocks 01*0.} 
\tiny
\begin{verbatim}    
(a) F, n=9   t  u   e  b      (b) F, n=9   t  u   e  b          (c) F, n=11    t  u   e  b
000010000    0  0  =1  1      101010101    0  0  =5  4          01010101010    0  0  =5  5
..11|....    1  1  =3  1      |.|.|10.|    1  1  =5  3          1|.|.|.|.|1    1  1  =7  4 
..|||11..    2  2  =5  1      |.01||..|    2  2  =5  2          ||10.|.|.||    2  2  =7  3
11|||||..    3  3  =7  1      |11|||..|    3  3  =7  1          |||11|.|.||    4  3  =9  2
|||||||11    4  4  =9  0      ||||||11|    4  4  =9  0          ||||||10.||    5  4  =9  1
             T  U                          T  U                 |||||||11||    7  5  =11 0
                                                                               T  U

(d) F, n=39                                t   u   e   b         (e) F, n=39                                t   u   e    b
000000000000000000010000000000000000000    0   0  =1   1         010101010101010101010101010101010101010    0   0  =19  19
.................11|...................    9   1  =3   1         .|.|.|.|.|.|.|.|.|.|.|.|.|10.|.|.|.|.|.    1   1  =19  18
.................|||11.................   23   2  =5   1         .|.|.|.|.|.|.|.|.|.|.|.|.||11|.|.|.|.|.    2   2  =21  17
...............11|||||.................   24   3  =7   1         .|.|.|.|.01|.|.|.|.|.|.|.|||||.|.|.|.|.    3   3  =21  16
.............11|||||||.................   27   4  =9   1         .|.|.|10..||.|.|.|.|.|.|.|||||.|.|.|.|.    4   4  =21  15
.............|||||||||11...............   37   5  =11  1         .|.|.||...||.|.|.|.|.|.01|||||.|.|.|.|.    5   5  =21  14
.............|||||||||00...............   41   6  =9   1         .|.|10|...||.|.|.|.|.|..||||||.|.|.|.|.    6   6  =21  14
...........11|||||||||.................   61   7  =11  1         .|.00.|...||.|.|.|.|.|..||||||.|.|.|.|.    7   7  =19  13
.........11|||||||||||.................   86   8  =13  1         .|..11|...||.|.|.|.|.|..||||||.|.|.|.|.    8   8  =21  13
.........|||||||||||00.................   91   9  =11  1         .|..|||...||.|.|.|.|.|..||||||.|10.|.|.    9   9  =21  12
.........00|||||||||...................   97  10  =9   1         .|..|||11.||.|.|.|.|.|..||||||.||..|.|.   11  10  =23  12
...........|||||||||11.................   99  11  =11  1         .|..|||||.||10.|.|.|.|..||||||.||..|.|.   12  11  =23  11
...........00|||||||||.................  102  12  =9   1         .|..|||||.|||..|.|10.|..||||||.||..|.|.   14  12  =23  10
.............|||||||||11...............  107  13  =11  1         .|..|||||.|||..|.||11|..||||||.||..|.|.   16  13  =25   9
.............|||||||||||11.............  115  14  =13  1         .|..|||||.|||.10.|||||..||||||.||..|.|.   19  14  =25   9
...........11|||||||||||||.............  117  15  =15  1         .|..|||||.|||.|..|||||..||||00.||..|.|.   20  15  =23   9
...........|||||||||||||||11...........  138  16  =17  1         .|..|||||.|||.|..|||||..||||...||.10.|.   21  16  =23   9
.........11|||||||||||||||||...........  142  17  =19  1         .|11|||||.|||.|..|||||..||||...||.|..|.   24  17  =25   8
.........00|||||||||||||||||...........  143  18  =17  1         .||||||||.|||.|..|||||..||||...||.|11|.   25  18  =27   7
.........11|||||||||||||||||...........  156  19  =19  1         .||||||||.|||.|..|||||..00||...||.||||.   26  19  =25   7
.......11|||||||||||||||||||...........  159  20  =21  1         .||||||||.|||.|..|||||....||...00.||||.   28  20  =23   6
.......|||||||||||||||||||00...........  171  21  =19  1         .||||||||.|||10..|||||....||......||||.   30  21  =23   5
.....11|||||||||||||||||||.............  185  22  =21  1         .||||||||10|||...|||||....||......||||.   33  22  =23   5
...11|||||||||||||||||||||.............  194  23  =23  1         .|||||||||.|00...|||||....||......||||.   37  23  =21   5
...|||||||||||||||||||||||11...........  208  24  =25  1         1|||||||||.|.....|||||....||......||||1   40  24  =23   4
...|||||||||||||||||||||||||11.........  210  25  =27  1         ||||||||||.|.....|||||11..||......|||||   42  25  =25   4
...|||||||||||||||||||||||||00.........  217  26  =25  1         ||||||||||.|.....|||||||..||....11|||||   52  26  =27   4
...00|||||||||||||||||||||||...........  224  27  =23  1         ||||||||||.|...11|||||||..||....|||||||   63  27  =29   4
.....|||||||||||||||||||||00...........  235  28  =21  1         ||||||||||.|...|||||||00..||....|||||||   65  28  =27   4
.....|||||||||||||||||||||11...........  239  29  =23  1         ||||||||||10...|||||||....||....|||||||   67  29  =27   3
.....|||||||||||||||||||||00...........  244  30  =21  1         |||||||||||....|||||00....||....|||||||   72  30  =25   3
.....|||||||||||||||||||||11...........  255  31  =23  1         |||||||||||..11|||||......||....|||||||   73  31  =27   3
.....|||||||||||||||||||||||11.........  257  32  =25  1         |||||||||||..|||||||......||..11|||||||   82  32  =29   3
.....|||||||||||||||||||||||||11.......  258  33  =27  1         |||||||||||..|||||||11....||..|||||||||   85  33  =31   3
.....|||||||||||||||||||||||||||11.....  302  34  =29  1         |||||||||||11|||||||||....||..|||||||||   88  34  =33   2
.....|||||||||||||||||||||||||||||11...  306  35  =31  1         ||||||||||||||||||||||..11||..|||||||||   89  35  =35   2
...11|||||||||||||||||||||||||||||||...  315  36  =33  1         ||||||||||||||||||||||11||||..|||||||||   97  36  =37   1
...00|||||||||||||||||||||||||||||||...  319  37  =31  1         ||||||||||||||||||||||||||00..|||||||||   98  37  =35   1
.....|||||||||||||||||||||||||||||00...  346  38  =29  1         ||||||||||||||||||||||||||11..|||||||||  107  38  =37   1
.....|||||||||||||||||||||||||||||11...  370  39  =31  1         ||||||||||||||||||||||||||||11|||||||||  118  39  =39   0
.....00|||||||||||||||||||||||||||||...  380  40  =29  1                                                    T   U
.....11|||||||||||||||||||||||||||||...  408  41  =31  1
.....|||||||||||||||||||||||||||||||11.  421  42  =33  1
...11|||||||||||||||||||||||||||||||||.  423  43  =35  1
1..|||||||||||||||||||||||||||||||||||1  428  44  =37  1
|11||||||||||||||||||||||||||||||||||||  439  45  =39  0
                                           T   U                                       
\end{verbatim}
\normalsize
\label{simruleF}
\end{table}   

\subsection{Rule HA}
\label{SimRuleHA}
\begin{table}[H]
\center
\caption{
Rule HA. Examples for initial configurations  (i) $10^{n-1}$ containing one \textit{1-block},   
or (ii) 0(01)* or 1(10)* containing the \textit{maximal number} of 1-blocks.
Only changed configurations are displayed. 
\textit{t}: micro-time step; \textit{u}: number of updates; \textit{e}: number of ones; 
\textit{b}: number of 1-blocks 01*0.} 
\tiny
\begin{verbatim}    
(a) HA,n=9  t  u   e  b       (b) HA,n=9   t  u   e  b          (c) HA,n=11   t  u   e  b 
000010000   0  0  =1  1       101010101    0  0  =5  4          01010101010   0  0  =5  5
..11|....   1  1  =3  1       |.01|.|.|    1  1  =5  3          .|.|.|.01|.   1  1  =4  4
11|||....   2  2  =5  1       |..||.01|    2  2  =5  2          .|10.|..||.   2  2  =3  3
|||||11..   3  3  =7  1       |11||..||    3  3  =7  1          1||..|..||1   3  3  =7  2
|||||||11   4  4  =9  1       |||||11||    4  4  =9  0          |||11|..|||   4  4  =9  1
            T  U                           T  U                 ||||||11|||   5  5  =11 0
                                                                              T  U                      

(d) HA, n=39                               t   u  e   b        (e) HA, n=39                               t   u  e    b
000000000000000000010000000000000000000    0  0  =1   1        010101010101010101010101010101010101010    0  0  =19  19
...................01..................   12  1  =1   1        .|.|.|10.|.|.|.|.|.|.|.|.|.|.|.|.|.|.|.    1  1  =19  18
....................01.................   14  2  =1   1        .|.|.||..|.|.|.|.|.|.|.|.|.|.01|.|.|.|.    2  2  =19  17
...................11|.................   19  3  =3   1        .|.|.||..|.|.|.|10.|.|.|.|.|..||.|.|.|.    3  3  =19  16
.................11|||.................   20  4  =5   1        .|10.||..|.|.|.||..|.|.|.|.|..||.|.|.|.    5  4  =19  15
...............11|||||.................   47  5  =7   1        .||11||..|.|.|.||..|.|.|.|.|..||.|.|.|.    6  5  =21  14
...............|||||||11...............   50  6  =9   1        .||||||..|.|.|.||..|.|.|.|.|..||10.|.|.    9  6  =21  13
...............|||||||00...............   52  7  =7   1        .||||||..|.|.|.||11|.|.|.|.|..|||..|.|.   10  7  =23  12
...............|||||||11...............   74  8  =9   1        .||||||..|.|.|.|||||.|10.|.|..|||..|.|.   12  8  =23  11
...............00|||||||...............   95  9  =7   1        .||||||..|.|.|.|||||.||..|.|11|||..|.|.   14  9  =25  10
.................|||||||11.............  102 10  =9   1        .||||||11|.|.|.|||||.||..|.||||||..|.|.   20 10  =27   9
.................00|||||||.............  109 11  =7   1        .|||||||||.|.|.|||||.||..|.||||||..01|.   21 11  =27   8
...................|||||||11...........  142 12  =9   1        .|||||||||.|.|.|||||.||..|.||||||.11||.   25 12  =29   8
.................11|||||||||...........  145 13  =11  1        .|||||||||.|.|.|||||.||..|10|||||.||||.   29 13  =29   8
.................|||||||||||11.........  155 14  =13  1        .|||||||||.|.|10||||.||..||.|||||.||||.   30 14  =29   8
...............11|||||||||||||.........  193 15  =15  1        .|||||||||.|.||.||||.||..||.|||00.||||.   32 15  =27   8
.............11|||||||||||||||.........  230 16  =17  1        .|||||||||.|.||.||||.||11||.|||...||||.   33 16  =29   7
.............00|||||||||||||||.........  245 17  =15  1        .|||||||||.|.||.||||.00||||.|||...||||.   34 17  =27   7
...............|||||||||||||00.........  250 18  =13  1        .|||||||||.|.||.||||...||||.|00...||||.   37 18  =25   7
...............|||||||||||00...........  265 19  =11  1        .|||||||||.|.||.||||...||||.|11...||||.   40 19  =27   7
.............11|||||||||||.............  273 20  =13  1        .|||||||||.|.||10|||...||||.|||...||||.   43 20  =27   7
...........11|||||||||||||.............  280 21  =15  1        .|||||||||.|.|||.|||11.||||.|||...||||.   46 21  =29   7
.........11|||||||||||||||.............  282 22  =17  1        .|||||||||.|10||.|||||.||||.|||...||||.   47 22  =29   7
.......11|||||||||||||||||.............  284 23  =19  1        .|||||||||.||.||.|||||.||||.|||11.||||.   50 23  =31   7
.......|||||||||||||||||||11...........  292 24  =21  1        .|||||||||.||10|.|||||.||||.|||||.||||.   51 24  =31   7
.....11|||||||||||||||||||||...........  294 25  =23  1        .|||||||||.|||.|.|||||.||||.00|||.||||.   53 25  =29   7
.....|||||||||||||||||||||||11.........  296 26  =25  1        .|||||||||.|||.01|||||.||||...|||.||||.   55 26  =29   6
...11|||||||||||||||||||||||||.........  304 27  =27  1        .|||||||||.|||..00||||.||||...|||.||||.   56 27  =27   6
.11|||||||||||||||||||||||||||.........  316 28  =29  1        .|||||||||10||....||||.||||...|||.||||.   57 28  =27   6
.|||||||||||||||||||||||||||||11.......  339 29  =31  1        .||||||||||.||....||||.||||.11|||.||||.   60 29  =29   6
.|||||||||||||||||||||||||||||||11.....  341 30  =33  1        .||||||||||.||....||||.||||10||||.||||.   62 30  =29   6
.|||||||||||||||||||||||||||||||00.....  344 31  =31  1        .||||||||||.||....||||.|||||10|||.||||.   63 31  =29   6
.|||||||||||||||||||||||||||||||11.....  345 32  =33  1        .||||||||||.||....||||.||||||10||.||||.   64 32  =29   6
1|||||||||||||||||||||||||||||||||....1  352 33  =35  1        .||||||||||10|....||||.|||||||.||.||||.   68 33  =29   6
||||||||||||||||||||||||||||||||||..11|  371 34  =37  1        .|||||||||||.|....||00.|||||||.||.||||.   69 34  =27   6
||||||||||||||||||||||||||||||||||11|||  425 35  =39  0        .|||||||||||.|....||...|||||||.00.||||.   70 35  =25   5
                                           T  U                .|||||||||||.|....||...|||||||11..||||.   73 36  =27   5
                                                               .00|||||||||.|....||...|||||||||..||||.   74 37  =25   5
                                                               ...|||||||||.|....||...|||||||||11||||.   77 38  =27   4
                                                               .11|||||||||.|....||...|||||||||||||||.   87 39  =29   4
                                                               .|||||||||||.|....||.11|||||||||||||||.   89 40  =31   4
                                                               .|||||||||||.01...||.|||||||||||||||||.   97 41  =31   4
                                                               1|||||||||||..|...||.|||||||||||||||||1   99 42  =33   3
                                                               ||||||||||||..|11.||.||||||||||||||||||  100 43  =35   3
                                                               ||||||||||||..|||.||10|||||||||||||||||  102 44  =35   3
                                                               ||||||||||||..|||.|00.|||||||||||||||||  105 45  =33   3
                                                               ||||||||||||11|||.|...|||||||||||||||||  106 46  =35   2
                                                               |||||||||||||||||.|.11|||||||||||||||||  121 47  =37   2
                                                               |||||||||||||||||.01|||||||||||||||||||  150 48  =37   1
                                                               |||||||||||||||||11||||||||||||||||||||  161 49  =39   0
                                                                                                          T  U
\end{verbatim}
\normalsize
\label{simruleHA}
\end{table}   

\subsection{Rule HB}
\label{SimRuleHB}
\begin{table}[H]
\vspace{-5mm}          
\center
\caption{
Rule HB. Examples for initial configurations  (i) $10^{n-1}$ containing one \textit{1-block},   
or (ii) 0(01)* or 1(10)* containing the \textit{maximal number} of 1-blocks.
\textit{t}: micro-time step; \textit{u}: number of updates; \textit{e}: number of ones; 
\textit{b}: number of 1-blocks 01*0.} 
\tiny
\begin{verbatim}    
(a) HB,n=9  t  u   e  b       (b) HB,n=9  t  u   e  b          (c) HB,n=11   t  u   e  b
000010000   0  0  =1  1       101010101   0  0  =5  4          01010101010   0  0  =5  5
....|11..   2  1  =3  1       0.|.|.|.0   2  1  =3  3          1|.|.|.|.|1   9  1  =7  4
..11|||..   3  2  =5  1       ..|.|.|11   3  2  =5  3          00.|.|.|.||  10  2  =5  4
11|||||..   4  3  =7  1       11|.|.|||   4  3  =7  2          ...|.|.|.00  12  3  =3  3
|||||||11   5  4  =9  0       |||.|.00|   8  4  =5  2          ...|.|.|11.  14  4  =5  3
            T  U              |||.|11.|   9  5  =7  2          1..|.|.|||1  17  5  =7  3
                              |||.00|.|  11  6  =5  2          |..|.|.00||  19  6  =5  3
                              0||...|.0  12  7  =3  2          0..|.|...|0  20  7  =3  3
                              .||.11|..  13  8  =5  2          .11|.|...|.  22  8  =5  3
                              .||.|||11  14  9  =7  2          .|00.|...|.  23  9  =3  3
                              .00.|||||  15 10  =5  1          .|...|11.|.  24 10  =5  3
                              11..|||||  18 11  =7  1          1|...|||.|1  26 11  =7  2
                              ||11|||||  19 12  =9  0          ||.11|||.||  27 12  =9  2
                                          T  U                 ||.|||||.00  29 13  =7  2
                                                               ||.|||00...  30 14  =5  2
                                                               ||.00|.....  31 15  =3  2
                                                               ||...|11...  34 16  =5  2
                                                               ||...|||11.  35 17  =7  2
                                                               00...|||||.  36 18  =5  1
                                                               1....|||||1  40 19  =7  1
                                                               |11..||||||  41 20  =9  1
                                                               |||11||||||  42 21  =11 0
                                                                            T  U
(d) HB, n=39                               t   u   e    b       (e) HB, n=39                               t   u   e   b
010101010101010101010101010101010101010    0   0  =19  19       000000000000000000010000000000000000000    0   0  =1   1
1|.|.|.|.|.|.|.|.|.|.|.|.|.|.|.|.|.|.|1   54   1  =21  18       ...................|11.................    4   1  =3   1
00.|.|.|.|.|.|.|.|.|.|.|.|.|.|.|.|.|.||   68   2  =19  18       ...................|||11...............    5   2  =5   1
...|.|.|.|.|.|.|.|.|.|.|.|.|.|.|.|.|.00   70   3  =17  17       ...................|||00...............    8   3  =3   1
.11|.|.|.|.|.|.|.|.|.|.|.|.|.|.|.|.|...   98   4  =19  17       ...................00|.................   16   4  =1   1
.|||.|.|.|.|.|.|.|.|.|.|.|.|.|.|.|.|11.  142   5  =21  17       ...................11|.................   35   5  =3   1
.|||.|.|.|.|.|.|.|.|.|.|.|.|.|.|.|.|00.  149   6  =19  17       ...................|||11...............   40   6  =5   1
.|||.|.|.|.|.|.|.|.|.|.|.|.|.|.|.|.|11.  159   7  =21  17       .................11|||||...............   42   7  =7   1
.|00.|.|.|.|.|.|.|.|.|.|.|.|.|.|.|.|||.  168   8  =19  17       .................|||||00...............   53   8  =5   1
.|.11|.|.|.|.|.|.|.|.|.|.|.|.|.|.|.|||.  169   9  =21  17       .................|||||11...............   55   9  =7   1
.|.00|.|.|.|.|.|.|.|.|.|.|.|.|.|.|.|||.  180  10  =19  17       ...............11|||||||...............   57  10  =9   1
.|...|.|.|.|.|.|.|.|.|.|.|.|.|.|.|.|00.  187  11  =17  17       .............11|||||||||...............   62  11  =11  1
.|11.|.|.|.|.|.|.|.|.|.|.|.|.|.|.|.|...  208  12  =19  17       .............|||||||||||11.............   63  12  =13  1
1|||.|.|.|.|.|.|.|.|.|.|.|.|.|.|.|.|..1  212  13  =21  17       .............|||||||||||||11...........   83  13  =15  1
||||.|.|.|.|.|.|.|.|.|.|.|.|.|.|.|.|11|  215  14  =23  16       .............00|||||||||||||...........   85  14  =13  1
||00.|.|.|.|.|.|.|.|.|.|.|.|.|.|.|.||||  217  15  =21  16       ...............|||||||||||00...........   93  15  =11  1
||11.|.|.|.|.|.|.|.|.|.|.|.|.|.|.|.||||  223  16  =23  16       ...............|||||||||00.............  112  16  =9   1
||00.|.|.|.|.|.|.|.|.|.|.|.|.|.|.|.||||  236  17  =21  16       ...............|||||||||11.............  115  17  =11  1
||11.|.|.|.|.|.|.|.|.|.|.|.|.|.|.|.||||  237  18  =23  16       .............11|||||||||||.............  132  18  =13  1
||00.|.|.|.|.|.|.|.|.|.|.|.|.|.|.|.||||  240  19  =21  16       .............|||||||||||||11...........  141  19  =15  1
||11.|.|.|.|.|.|.|.|.|.|.|.|.|.|.|.||||  241  20  =23  16       .............|||||||||||||||11.........  147  20  =17  1
||||.|.|.|.|.|.|.|.|.|.|.|.|.|.|.|.00||  247  21  =21  16       .............|||||||||||||||||11.......  154  21  =19  1
=======================================                         ...........11|||||||||||||||||||.......  156  22  =21  1
|.||00...|||||.|.|.|||||.|||||.|.|||||| 2080 500  =27   8       ...........|||||||||||||||||||||11.....  157  23  =23  1
|.||11...|||||.|.|.|||||.|||||.|.|||||| 2081 501  =29   8       ...........|||||||||||||||||||||||11...  168  24  =25  1
0.||||...|||||.|.|.|||||.|||||.|.|||||0 2082 502  =27   8       ...........|||||||||||||||||||||||||11.  176  25  =27  1
..||||...00|||.|.|.|||||.|||||.|.|||||. 2088 503  =25   8       .........11|||||||||||||||||||||||||||.  178  26  =29  1
1.||||.....|||.|.|.|||||.|||||.|.|||||1 2090 504  =27   8       1........|||||||||||||||||||||||||||||1  209  27  =31  1
|.||00.....|||.|.|.|||||.|||||.|.|||||| 2096 505  =25   8       |11......||||||||||||||||||||||||||||||  214  28  =33  1
|.00.......|||.|.|.|||||.|||||.|.|||||| 2100 506  =23   7       |||....11||||||||||||||||||||||||||||||  215  29  =35  1
|..........|||.|.|.00|||.|||||.|.|||||| 2110 507  =21   7       |||..11||||||||||||||||||||||||||||||||  223  30  =37  1
|..........|00.|.|...|||.|||||.|.|||||| 2114 508  =19   7       |||11||||||||||||||||||||||||||||||||||  238  31  =39  0
|..........|...|.|.11|||.|||||.|.|||||| 2116 509  =21   7                                                  T   U	   
|11........|...|.|.|||||.|||||.|.|||||| 2118 510  =23   7
|||........|...|.|.|||||.00|||.|.|||||| 2120 511  =21   7
|||........|...|.|.|||||...|||.|.00|||| 2121 512  =19   7
|||........|...|.|.|||||...|00.|...|||| 2123 513  =17   7
|||........|...|.|.00|||...|...|...|||| 2127 514  =15   7
|||......11|...|.|...|||...|...|...|||| 2130 515  =17   7
|||......|||...|.|...|||...|...|11.|||| 2132 516  =19   7
|||......|||...|.|...|||...|...|||.00|| 2133 517  =17   7
|||......|||...|.|...|||...|.11|||...|| 2134 518  =19   7
|||......|||...|.|...|||...|.|||00...|| 2135 519  =17   7
|||......|||...|.|...00|...|.|||.....|| 2136 520  =15   7
=======================================					 		
|||||||||||||||||||||||......00|||||||| 4381 970  =31   1
|||||||||||||||||||||||11......|||||||| 4383 971  =33   1
|||||||||||||||||||||||||11....|||||||| 4393 972  =35   1
|||||||||||||||||||||||||||..11|||||||| 4397 973  =37   1
|||||||||||||||||||||||||00..|||||||||| 4399 974  =35   1
|||||||||||||||||||||||||....00|||||||| 4402 975  =33   1
|||||||||||||||||||||||||11....|||||||| 4423 976  =35   1
|||||||||||||||||||||||||||11..|||||||| 4437 977  =37   1
|||||||||||||||||||||||||||||..00|||||| 4452 978  =35   1
|||||||||||||||||||||||||||||11..|||||| 4480 979  =37   1
|||||||||||||||||||||||||||||||..00|||| 4489 980  =35   1
|||||||||||||||||||||||||||||00....|||| 4505 981  =33   1
|||||||||||||||||||||||||||||11....|||| 4520 982  =35   1
|||||||||||||||||||||||||||||||11..|||| 4525 983  =37   1
|||||||||||||||||||||||||||||||00..|||| 4532 984  =35   1
|||||||||||||||||||||||||||||||..11|||| 4533 985  =37   1
|||||||||||||||||||||||||||||00..|||||| 4549 986  =35   1
|||||||||||||||||||||||||||||11..|||||| 4566 987  =37   1
|||||||||||||||||||||||||||||||11|||||| 4568 988  =39   0
                                           T   U
\end{verbatim}
\normalsize
\label{simruleHB}
\end{table}   
\subsection{Rule HC}
\label{SimRuleHC}
\begin{table}[H]
\center
\caption{
Rule HC. Examples for initial configurations  (i) $10^{n-1}$ containing one \textit{1-block},   
or (ii) 0(01)* or 1(10)* containing the \textit{maximal number} of 1-blocks.
Only changed configurations are displayed. 
\textit{t}: micro-time step; \textit{u}: number of updates; \textit{e}: number of ones; 
\textit{b}: number of 1-blocks 01*0.} 
\tiny
\begin{verbatim}    
(a) HC,n=9  t  u   e  b       (b) HC,n=9  t  u   e  b          (c) HC,n=11   t  u   e  b
000010000   0  0  =1  1       101010101   0  0  =5  4          01010101010   0  0  =5  5
....|11..   3  1  =3  1       0.|.|.|.0   1  1  =3  3          1|.|.|.|.|1   1  1  =7  4
....|||11   4  2  =5  1       ..|.|.|11   5  2  =5  3          ||.|.|.|.00   3  2  =5  4
11..|||||   5  3  =7  1       ..|.|.00|   6  3  =3  3          00.|.|.|...   4  3  =3  3
||11|||||   6  4  =9  0       ..|.|11.|  12  4  =5  3          ...|.|.|11.   6  4  =5  3
            T  U              11|.|||.|  13  5  =7  2          1..|.|.|||1   7  5  =7  3
                              0||.|||.0  16  6  =5  2          |..|.|.00||   8  6  =5  3
                              .00.|||..  18  7  =3  1          |..|.|11.||   9  7  =7  3
                              ....00|..  19  8  =1  1          |..|.|||.00  10  8  =5  3
                              ......|11  20  9  =3  1          |11|.|||...  11  9  =7  2
                              11....|||  21 10  =5  1          00||.|||...  22 10  =5  2
                              ||11..|||  22 11  =7  1          ..00.|||...  23 11  =3  1
                              ||||11|||  24 12  =9  0          .....|||11.  31 12  =5  1
                                          T  U                 1....|||||1  44 13  =7  1
                                                               |11..||||||  51 14  =9  1
                                                               |||11||||||  52 15  =11 0
                                                                             T  U
                                                               
(d) HC, n=39                               t   u   e    b        (e) HC, n=39                               t   u   e   b
010101010101010101010101010101010101010    0    0  =19  19       000000000000000000010000000000000000000    0   0  =1   1
1|.|.|.|.|.|.|.|.|.|.|.|.|.|.|.|.|.|.|1   39    1  =21  18       ...................|11.................   11   1  =3   1
||.|.|.|.|.|.|.|.|.|.|.|.|.|.|.|.|.|.00   52    2  =19  18       ...................00|.................   63   2  =1   1
||.|.|.|.|.|.|.|.|.|.|.|.|.|.|.|.|.|11.   62    3  =21  18       .....................|11...............   82   3  =3   1
||.|.|.|.|.|.|.|.|.|.|.|.|.|.|.|.|.00|.  113    4  =19  18       .....................|||11.............   86   4  =5   1
||.|.|.|.|.|.|.|.|.|.|.|.|.|.|.|.|11.|.  132    5  =21  18       .....................|||||11...........  140   5  =7   1
00.|.|.|.|.|.|.|.|.|.|.|.|.|.|.|.|||.|.  138    6  =19  17       .....................00|||||...........  143   6  =5   1
...|.|.|.|.|.|.|.|.|.|.|.|.|.|.|.00|.|.  149    7  =17  17       .......................|||||11.........  154   7  =7   1
1..|.|.|.|.|.|.|.|.|.|.|.|.|.|.|...|.|1  199    8  =19  17       .......................|||||||11.......  156   8  =9   1
|..|.|.|.|.|.|.|.|.|.|.|.|.|.|.|11.|.||  200    9  =21  17       .......................|||||||||11.....  169   9  =11  1
|11|.|.|.|.|.|.|.|.|.|.|.|.|.|.|||.|.||  207   10  =23  16       .......................|||||||||||11...  172  10  =13  1
||||.|.|.|.|.|.|.|.|.|.|.|.|.|.00|.|.||  277   11  =21  16       .......................|||||||||||||11.  179  11  =15  1
||||.|.|.|.|.|.|.|.|.|.|.|.|.|...|.|.00  291   12  =19  16       1......................|||||||||||||||1  197  12  =17  1
00||.|.|.|.|.|.|.|.|.|.|.|.|.|...|.|...  293   13  =17  16       |11....................||||||||||||||||  233  13  =19  1
..||.|.|.|.|.|.|.|.|.|.|.|.|.|11.|.|...  299   14  =19  16       |||....................00||||||||||||||  237  14  =17  1
..||.|.|.|.|.|.|.|.|.|.|.|.|.|||.|.|11.  302   15  =21  16       |||11....................||||||||||||||  248  15  =19  1
1.||.|.|.|.|.|.|.|.|.|.|.|.|.|||.|.|||1  305   16  =23  16       |||||11..................||||||||||||||  272  16  =21  1
|.||.|.|.|.|.|.|.|.|.|.|.|.|.|||.|.00||  312   17  =21  16       |||||||11................||||||||||||||  308  17  =23  1
|.||.|.|.|.|.|.|.|.|.|.|.|.|.|||.|...00  323   18  =19  16       |||||||||11..............||||||||||||||  334  18  =25  1
|.00.|.|.|.|.|.|.|.|.|.|.|.|.|||.|.....  341   19  =17  15       |||||||||||11............||||||||||||||  351  19  =27  1
|....|.|.|.|.|.|.|.|.|.|.|.|.00|.|.....  353   20  =15  15       |||||||||||||11..........||||||||||||||  359  20  =29  1
==========================================================       |||||||||||||||..........00||||||||||||  365  21  =27  1
..|||.....00||||||.|||.....|.|...|..... 1419  170  =15   6       |||||||||||||||11..........||||||||||||  375  22  =29  1
..00|.......||||||.|||.....|.|...|..... 1437  171  =13   6       |||||||||||||||||11........||||||||||||  393  23  =31  1
....|.......00||||.|||.....|.|...|..... 1440  172  =11   6       |||||||||||||||||||11......||||||||||||  422  24  =33  1
....|.........00||.|||.....|.|...|..... 1442  173  =9    6       |||||||||||||||||||||11....||||||||||||  434  25  =35  1
....|...........||.|||.....|.|...|11... 1444  174  =11   6       |||||||||||||||||||||||11..||||||||||||  462  26  =37  1
....|...........||.|||11...|.|...|||... 1446  175  =13   6       |||||||||||||||||||||||||11||||||||||||  480  27  =39  0
....|...........||.|||||11.|.|...|||... 1449  176  =15   6                                                  T   U
....|...........||.00|||||.|.|...|||... 1452  177  =13   6       
....|...........||...00|||.|.|...|||... 1453  178  =11   6       
....|...........00.....|||.|.|...|||... 1459  179  =9    5       
....|..................00|.|.|...|||... 1478  180  =7    5       
....|11..................|.|.|...|||... 1480  181  =9    5
....|||..................|.|.|...|||11. 1496  182  =11   5
1...|||..................|.|.|...|||||1 1499  183  =13   5
|11.|||..................|.|.|...|||||| 1501  184  =15   5
|||.|||..................|.|.|11.|||||| 1505  185  =17   5
|||.|||..................|.|.00|.|||||| 1541  186  =15   5
|||.|||..................|.|11.|.|||||| 1543  187  =17   5
|||.00|..................|.|||.|.|||||| 1550  188  =15   5
|||...|11................|.|||.|.|||||| 1554  189  =17   5
|||...|||................|.|||.|.00|||| 1562  190  =15   5
==========================================================
|||||||......00|||||||||||||||||||||||| 3787  350  =31   1
|||||||11......|||||||||||||||||||||||| 3817  351  =33   1
|||||||||11....|||||||||||||||||||||||| 3818  352  =35   1
|||||||||||....00|||||||||||||||||||||| 3838  353  =33   1
|||||||||||11....|||||||||||||||||||||| 3842  354  =35   1
|||||||||||||11..|||||||||||||||||||||| 3859  355  =37   1
|||||||||||||||..00|||||||||||||||||||| 3860  356  =35   1
|||||||||||||||11..|||||||||||||||||||| 3869  357  =37   1
|||||||||||||||||..00|||||||||||||||||| 3876  358  =35   1
|||||||||||||||||....00|||||||||||||||| 3883  359  =33   1
|||||||||||||||||11....|||||||||||||||| 3891  360  =35   1
|||||||||||||||||||11..|||||||||||||||| 3902  361  =37   1
|||||||||||||||||||||..00|||||||||||||| 3909  362  =35   1
|||||||||||||||||||||....00|||||||||||| 3930  363  =33   1
|||||||||||||||||||||11....|||||||||||| 3962  364  =35   1
|||||||||||||||||||||||11..|||||||||||| 3992  365  =37   1
|||||||||||||||||||||||||..00|||||||||| 4001  366  =35   1
|||||||||||||||||||||||||....00|||||||| 4002  367  =33   1
|||||||||||||||||||||||||......00|||||| 4045  368  =31   1
|||||||||||||||||||||||||11......|||||| 4054  369  =33   1
|||||||||||||||||||||||||||11....|||||| 4086  370  =35   1
|||||||||||||||||||||||||||||11..|||||| 4102  371  =37   1
|||||||||||||||||||||||||||||||11|||||| 4118  372  =39   0																																			
                                           T    U																																																																	
\end{verbatim}
\normalsize
\label{simruleHC}
\end{table}   

\subsection{Rule HD}
\label{SimRuleHD}

\begin{table}[H]
\center
\caption{
Rule HD. Examples for initial configurations  (i) $10^{n-1}$ containing one \textit{1-block},   
or (ii) 0(01)* or 1(10)* containing the \textit{maximal number} of 1-blocks.
Only changed configurations are displayed. 
\textit{t}: micro-time step; \textit{u}: number of updates; \textit{e}: number of ones; 
\textit{b}: number of 1-blocks 01*0.} 
\tiny
\begin{verbatim}    
(a) HD,n=9  t  u   e  b       (b) HD,n=9  t  u   e  b           (c) HD,n=11   t  u   e   b
000010000   0  0  =1  1       101010101   0  0  =5  4           01010101010   0  0  =5   5
....|11..   1  1  =3  1       |1|1|.|.|   1  1  =7  2           .|.|1|1|.|.   1  1  =7   3
..11|||..   2  2  =5  1       |||||1|1|   3  2  =9  0           1|1|||||.|.   2  2  =9   2
..|||||11   3  3  =7  1                   T  U                  ||||||||1|1   3  3  =11  0
11|||||||   4  4  =9  0                                                       T  U
            T  U                                                
                                                                                                                              
(d) HD, n=39                               t   u   e    b        (e) HD, n=39                               t   u    e   b
010101010101010101010101010101010101010    0   0   =19  19       000000000000000000010000000000000000000    0   0   =1   1
.|.|.|.|.|.|.|.|.|.|.0.0.|.|.|.|.|.|.|.    1   1   =17  17       .................11|...................    4   4   =3   1
.|.|1|1|.|.|.|.|.|.|.....|.|.|.|.|.|.|.    2   2   =19  15       .................|00...................   11   2   =1   1
.|.|||||.|.|.|.0.0.|.....|.|.|.|.|.|.|.    3   3   =17  13       ...............11|.....................   30   3   =3   1
.|.|||00.|.|.|.....|.....|.|.|.|.|.|.|.    5   4   =15  13       ...............00|.....................   31   4   =1   1
.|.|||11.|.|.|.....|.....|.|.|.|.|.|.|.    6   5   =17  13       ................1|1....................   46   5   =3   1
.|.|||||.|.|.|.....|....1|1|.|.|.|.|.|.    7   6   =19  12       ..............11|||....................   62   6   =5   1
.|.|||||1|1|.|.....|....||||.|.|.|.|.|.    8   7   =21  10       ..............00|||....................   77   7   =3   1
.|.|||||||||.|.....|....||||.|.|1|1|.|.    9   8   =23   8       ................00|....................   85   8   =1   1
1|.|||||||||.|.....|....||||.|.|||||.|1   11   9   =25   7       .................1|1...................   89   9   =3   1
||.|||||||||.|.....|..11||||.|.|||||.||   12  10   =27   7       .................|||11.................   90  10   =5   1
||.|||||||||.|...11|..||||||.|.|||||.||   13  11   =29   7       .................|||||11...............   94  11   =7   1
||.|||||||||1|1..|||..||||||.|.|||||.||   14  12   =31   6       .................|||||||11.............  113  12   =9   1
||.||||||||||||..|||..||||||.|.||||0.0|   19  13   =29   6       .................|||||||||11...........  125  13   =11  1
||.||||||||||||11|||..||||||.|.||||...|   21  14   =31   5       ...............11|||||||||||...........  137  14   =13  1
||.|||||||||||||||||11||||||.|.||||...|   23  15   =33   4       ...............|||||||||||00...........  141  15   =11  1
||.|||||||||||||||||||||||00.|.||||...|   27  16   =31   4       .............11|||||||||||.............  142  16   =13  1
||.|||||||||||||||||||||||..1|1||||...|   29  17   =33   3       ...........11|||||||||||||.............  157  17   =15  1
0|.|||||||||||||||||||||||..|||||||...0   31  18   =31   3       .........11|||||||||||||||.............  160  18   =17  1
.0.0||||||||||||||||||||||..|||||||....   32  19   =29   2       .......11|||||||||||||||||.............  163  19   =19  1
....||||||||||||||||||||||..|||||||11..   33  20   =31   2       .......|||||||||||||||||00.............  164  20   =17  1
....||||||||||||||||||||||11|||||||||..   34  21   =33   1       .....11|||||||||||||||||...............  167  21   =19  1
....|||||||||||||||||||||||||||||||00..   37  22   =31   1       .....|||||||||||||||||||11.............  170  22   =21  1
..11|||||||||||||||||||||||||||||||....   46  23   =33   1       .....00|||||||||||||||||||.............  185  23   =19  1
..|||||||||||||||||||||||||||||||||11..   54  24   =35   1       .......|||||||||||||||||||11...........  194  24   =21  1
11|||||||||||||||||||||||||||||||||||..   55  25   =37   1       .....11|||||||||||||||||||||...........  196  25   =23  1
|||||||||||||||||||||||||||||||||||||11   58  26   =39   0       ...11|||||||||||||||||||||||...........  209  26   =25  1
                                           T   U                 ...|||||||||||||||||||||||||11.........  224  27   =27  1
                                                                 .11|||||||||||||||||||||||||||.........  225  28   =29  1
                                                                 1|||||||||||||||||||||||||||||........1  230  29   =31  1
                                                                 0|||||||||||||||||||||||||||||........0  245  30   =29  1
                                                                 1|||||||||||||||||||||||||||||........1  249  31   =31  1
                                                                 ||||||||||||||||||||||||||||||11......|  257  32   =33  1
                                                                 ||||||||||||||||||||||||||||||||11....|  272  33   =35  1
                                                                 ||||||||||||||||||||||||||||||||||11..|  298  34   =37  1
                                                                 ||||||||||||||||||||||||||||||||||||11|  300  35   =39  0
                                                                                                            T   U
\end{verbatim}
\normalsize
\label{simruleHD}
\end{table}   

\newpage
\section{APPENDIX 2: Transition matrices for $n=5,7, 9$}


\subsection{Rule F}
\footnotesize
Rule \textit{F} is neighborhood-symmetric. 
The transition matrices are given in Table 
\ref{TMRuleF}. 
The indexes of the matrices (I, II, IV) represent state equivalents under shift,
and the indexes of (I, III, V)              
represent state equiv. under shift and mirror. 
The $(i,j)$ entries are the probabilities $p(i,j)$ to be divided by $n$ to change from state $i$ to $j$. 
\normalsize

\begin{table}[H]

\center
\caption{
Rule \textit{F} transition matrices for $n=5, 7, 9$. A dot represents a zero.
(II) The state 13 (the mirror of 11) is included in the matrix, and
(III) the state 13 is excluded.
(IV) The 29 indexes of the matrix are state representatives of equivalents under shift. 
(V) The 22 indexes are representatives of state equivalents under shift  and mirror.
}
\tiny
\begin{verbatim}        
(I) (n = 5)                  (II) (n = 7, shift representatives only)          (III) (n = 7, without 13, is mirror of 11
             1  7  11 31                    1  7  11 13 19 21 31 47 55 127            1  7  11 19 21 31 47 55 12      
            ------------                    ------------------------------           ---------------------------
00001 =  1   3  2  .  .       0000001 =  1  5  2  .  .  .  .  .  .  .  .           1  5  2  .  .  .  .  .  .  .
00111 =  7   1  2  .  2       0000111 =  7  2  3  .  .  .  .  2  .  .  .           7  2  3  .  .  .  2  .  .  .
01011 = 11   1  2  2  .       0001011 = 11  1  1  1  1  1  .  .  1  1  .          11  1  1  2  1  .  .  1  1  .
11111 = 31   .  .  .  5       0001101 = 13  1  1  1  1  1  .  .  1  1  .          19  1  .  2  2  .  2  .  .  .
            ------------      0010011 = 19  1  .  1  1  2  .  2  .  .  .          21  .  .  2  2  2  .  1  .  .
                              0010101 = 21  .  .  1  1  2  2  .  1  .  .          31  .  2  .  .  .  4  .  .  1
                              0011111 = 31  .  2  .  .  .  .  4  .  .  1          47  .  .  2  .  .  2  1  2  .
                              0101111 = 47  .  .  1  1  .  .  2  1  2  .          55  .  1  2  .  .  .  2  2  .
                              0110111 = 55  .  1  1  1  .  .  .  2  2  .         127  .  .  .  .  .  .  .  .  7
                              1111111 =127  .  .  .  .  .  .  .  .  .  7             ----------------------------
                                           ------------------------------

 (IV) (n = 9, shift representatives)                                107   117   171   223   511 
     1  7  11 13 19 21 25 31 35 37 41 47 55 59 61 73 79 87 91 93 103   109   127   191   239
    ------------------------------------------------------------------------------------------
  1  7  2  .  .  .  .  .  .  .  .  .  .  .  .  .  .  .  .  .  .  .  .  .  .  .  .  .  .  .  .
  7  2  5  .  .  .  .  .  2  .  .  .  .  .  .  .  .  .  .  .  .  .  .  .  .  .  .  .  .  .  .
 11  1  1  3  1  1  .  .  .  .  .  .  1  .  1  .  .  .  .  .  .  .  .  .  .  .  .  .  .  .  .
 13  1  1  1  3  .  .  1  .  .  .  .  .  1  .  1  .  .  .  .  .  .  .  .  .  .  .  .  .  .  .
 19  1  .  1  .  3  .  .  1  1  .  .  .  .  .  .  .  1  .  .  .  1  .  .  .  .  .  .  .  .  .
 21  .  .  1  1  1  1  1  .  .  1  1  .  .  .  .  .  .  1  .  .  .  .  .  1  .  .  .  .  .  .
 25  1  .  .  1  .  .  3  1  1  .  .  .  .  .  .  .  1  .  .  .  1  .  .  .  .  .  .  .  .  .
 31  .  2  .  .  .  .  .  5  .  .  .  .  .  .  .  .  .  .  .  .  .  .  .  .  2  .  .  .  .  .
 35  1  .  .  .  1  .  1  .  2  .  .  1  1  1  1  .  .  .  .  .  .  .  .  .  .  .  .  .  .  .
 37  .  .  .  .  1  1  .  .  1  .  2  .  .  .  1  1  .  1  .  1  .  .  .  .  .  .  .  .  .  .
 41  .  .  .  .  .  1  1  .  1  2  .  1  .  .  .  1  .  .  .  1  .  .  .  1  .  .  .  .  .  .
 47  .  .  1  .  .  .  .  1  1  .  .  2  1  .  .  .  1  .  .  .  .  .  .  .  .  .  1  .  1  .
 55  .  1  .  1  .  .  .  .  1  .  .  1  2  1  .  .  .  .  .  .  .  .  .  .  .  .  .  1  1  .
 59  .  1  1  .  .  .  .  .  1  .  .  .  1  2  1  .  .  .  .  .  .  .  .  .  .  .  .  1  1  .
 61  .  .  .  1  .  .  .  1  1  .  .  .  .  1  2  .  1  .  .  .  .  .  .  .  .  .  1  .  1  .
 73  .  .  .  .  .  .  .  .  .  3  3  .  .  .  .  .  3  .  .  .  .  .  .  .  .  .  .  .  .  .
 79  .  .  .  .  1  .  1  .  .  .  .  1  .  .  1  .  3  .  .  .  .  .  .  .  2  .  .  .  .  .
 87  .  .  .  .  .  1  .  .  .  1  .  .  1  .  .  .  1  1  1  1  1  .  .  .  .  .  1  .  .  .
 91  .  .  1  .  .  .  1  .  .  .  .  .  .  1  .  .  .  1  1  1  .  1  1  .  .  .  .  1  .  .
 93  .  .  .  .  .  .  .  .  .  1  1  1  .  .  1  .  .  1  1  .  .  .  1  1  .  .  .  .  1  .
103  .  1  .  .  1  .  1  .  .  .  .  .  .  .  .  .  .  .  .  .  4  .  .  .  2  .  .  .  .  .
107  .  .  1  1  .  .  .  .  .  .  .  .  .  .  .  .  .  .  1  .  2  2  1  .  .  .  1  .  .  .
109  .  .  .  1  1  .  .  .  .  .  .  .  1  .  .  .  .  .  1  1  .  1  1  1  .  .  .  1  .  .
117  .  .  .  .  .  1  .  .  .  .  1  .  .  1  .  .  1  .  .  1  1  .  1  1  .  .  1  .  .  .
127  .  .  .  .  .  .  .  2  .  .  .  .  .  .  .  .  .  .  .  .  .  .  .  .  6  .  .  .  .  1
171  .  .  .  .  .  1  .  .  .  .  .  .  .  .  .  .  .  1  1  .  .  2  1  1  .  2  .  .  .  .
191  .  .  .  .  .  .  .  .  .  .  .  1  .  .  1  .  .  .  .  .  .  .  .  .  2  .  3  2  .  .
223  .  .  .  .  .  .  .  1  .  .  .  .  1  1  .  .  .  .  .  .  .  .  .  .  .  .  2  2  2  .
239  .  .  .  .  .  .  .  .  .  .  .  1  1  1  1  .  .  .  .  .  .  .  .  .  .  .  .  2  3  .
511  .  .  .  .  .  .  .  .  .  .  .  .  .  .  .  .  .  .  .  .  .  .  .  .  .  .  .  .  .  9                 
    ------------------------------------------------------------------------------------------     
(V) (n = 9, shift & mirror representatives)          107   171   223   511
     1  7  11 19 21 31 35 37 47 55 73 79 87 91 93 103   127   191   239
    ----------------------------------------------------------------------
  1  7  2  .  .  .  .  .  .  .  .  .  .  .  .  .  .  .  .  .  .  .  .  .
  7  2  5  .  .  .  2  .  .  .  .  .  .  .  .  .  .  .  .  .  .  .  .  .
 11  1  1  4  1  .  .  .  .  1  1  .  .  .  .  .  .  .  .  .  .  .  .  .
 19  1  .  1  3  .  1  1  .  .  .  .  1  .  .  .  1  .  .  .  .  .  .  .
 21  .  .  2  2  1  .  .  2  .  .  .  .  2  .  .  .  .  .  .  .  .  .  .
 31  .  2  .  .  .  5  .  .  .  .  .  .  .  .  .  .  .  2  .  .  .  .  .
 35  1  .  .  2  .  .  2  .  2  2  .  .  .  .  .  .  .  .  .  .  .  .  .
 37  .  .  .  1  1  .  1  2  1  .  1  .  1  .  1  .  .  .  .  .  .  .  .
 47  .  .  1  .  .  1  1  .  2  1  .  1  .  .  .  .  .  .  .  1  .  1  .
 55  .  1  1  .  .  .  1  .  1  3  .  .  .  .  .  .  .  .  .  .  1  1  .
 73  .  .  .  .  .  .  .  6  .  .  .  3  .  .  .  .  .  .  .  .  .  .  .
 79  .  .  .  2  .  .  .  .  2  .  .  3  .  .  .  .  .  2  .  .  .  .  .
 87  .  .  .  .  1  .  .  1  .  1  .  1  1  1  1  1  .  .  .  1  .  .  .
 91  .  .  1  1  .  .  .  .  .  1  .  .  1  2  1  .  1  .  .  .  1  .  .
 93  .  .  .  .  .  .  .  2  2  .  .  .  2  2  .  .  .  .  .  .  .  1  .
103  .  1  .  2  .  .  .  .  .  .  .  .  .  .  .  4  .  2  .  .  .  .  .
107  .  .  2  .  .  .  .  .  .  .  .  .  .  2  .  2  2  .  .  1  .  .  .
127  .  .  .  .  .  2  .  .  .  .  .  .  .  .  .  .  .  6  .  .  .  .  1
171  .  .  .  .  1  .  .  .  .  .  .  .  2  2  .  .  2  .  2  .  .  .  .
191  .  .  .  .  .  .  .  .  2  .  .  .  .  .  .  .  .  2  .  3  2  .  .
223  .  .  .  .  .  1  .  .  .  2  .  .  .  .  .  .  .  .  .  2  2  2  .
239  .  .  .  .  .  .  .  .  2  2  .  .  .  .  .  .  .  .  .  .  2  3  .
511  .  .  .  .  .  .  .  .  .  .  .  .  .  .  .  .  .  .  .  .  .  .  9 
    ---------------------------------------------------------------------      
\end{verbatim}
\normalsize
\label{TMRuleF}
\end{table}   

\newpage
\subsection{Rule HA}

Rule \textit{HA} is not neighborhood-symmetric. The transition matrices are given in Table 
\ref{TMRuleHA}. The indexes
of the matrices represent state equivalents under shift. The $(i, j)$ entries are the probabilities
$p(i, j)$ to be divided by $n$ to change from state $i$ to $j$.

\begin{table}[H]
\center
\caption{
Rule \textit{HA} transition matrices for $n=5, 7, 9$. 
 The indexes of the matrix are state representatives of equivalents under shift. 
}
\tiny
\begin{verbatim}
(I) (n = 5)                     (II) (n = 7, shift representatives)
             1  7  11 31                      1  7  11 13 19 21 31 47 55 127      
            ------------                      -------------------------------    
00001 =  1   3  2  0  0         0000001 =  1  5  2  .  .  .  .  .  .  .  .     
00111 =  7   1  2  0  2         0000111 =  7  2  3  .  .  .  .  2  .  .  .     
01011 = 11   1  2  2  0         0001011 = 11  1  1  2  1  .  .  .  1  1  .     
11111 = 31   0  0  0  5         0001101 = 13  1  1  .  2  1  .  .  1  1  .     
            ------------        0010011 = 19  1  .  1  .  3  .  2  .  .  .     
                                0010101 = 21  .  .  1  1  2  2  .  1  .  .     
                                0011111 = 31  .  2  .  .  .  .  4  .  .  1     
                                0101111 = 47  .  .  1  1  .  .  2  2  1  .     
                                0110111 = 55  .  1  1  1  .  .  .  1  3  .     
                                1111111 =127  .  .  .  .  .  .  .  .  .  7     
                                             -------------------------------              
                                        
(III) (n = 9, shift representatives)                                107   117   171   223   511
     1  7  11 13 19 21 25 31 35 37 41 47 55 59 61 73 79 87 91 93 103   109   127   191   239
    -------------------------------------------------------------------------------------------  
  1  7  2  .  .  .  .  .  .  .  .  .  .  .  .  .  .  .  .  .  .  .  .  .  .  .  .  .  .  .  .
  7  2  5  .  .  .  .  .  2  .  .  .  .  .  .  .  .  .  .  .  .  .  .  .  .  .  .  .  .  .  .
 11  1  1  4  1  .  .  .  .  .  .  .  1  .  1  .  .  .  .  .  .  .  .  .  .  .  .  .  .  .  .
 13  1  1  .  4  .  .  1  .  .  .  .  .  1  .  1  .  .  .  .  .  .  .  .  .  .  .  .  .  .  .
 19  1  .  1  .  4  .  .  1  .  .  .  .  .  .  .  .  1  .  .  .  1  .  .  .  .  .  .  .  .  .
 21  .  .  1  1  1  2  1  .  .  .  1  .  .  .  .  .  .  1  .  .  .  .  .  1  .  .  .  .  .  .
 25  1  .  .  .  .  .  4  1  1  .  .  .  .  .  .  .  1  .  .  .  1  .  .  .  .  .  .  .  .  .
 31  .  2  .  .  .  .  .  5  .  .  .  .  .  .  .  .  .  .  .  .  .  .  .  .  2  .  .  .  .  .
 35  1  .  .  .  1  .  .  .  3  .  .  1  1  1  1  .  .  .  .  .  .  .  .  .  .  .  .  .  .  .
 37  .  .  .  .  1  1  .  .  1  2  .  .  .  .  1  1  .  1  .  1  .  .  .  .  .  .  .  .  .  .
 41  .  .  .  .  .  .  1  .  1  2  2  1  .  .  .  .  .  .  .  1  .  .  .  1  .  .  .  .  .  .
 47  .  .  1  .  .  .  .  1  1  .  .  3  1  .  .  .  .  .  .  .  .  .  .  .  .  .  1  .  1  .
 55  .  1  .  1  .  .  .  .  1  .  .  .  3  1  .  .  .  .  .  .  .  .  .  .  .  .  .  1  1  .
 59  .  1  1  .  .  .  .  .  1  .  .  .  .  3  1  .  .  .  .  .  .  .  .  .  .  .  .  1  1  .
 61  .  .  .  1  .  .  .  1  1  .  .  .  .  .  3  .  1  .  .  .  .  .  .  .  .  .  1  .  1  .
 73  .  .  .  .  .  .  .  .  .  .  3  .  .  .  .  3  3  .  .  .  .  .  .  .  .  .  .  .  .  .
 79  .  .  .  .  1  .  1  .  .  .  .  1  .  .  .  .  4  .  .  .  .  .  .  .  2  .  .  .  .  .
 87  .  .  .  .  .  1  .  .  .  1  .  .  1  .  .  .  1  2  1  .  1  .  .  .  .  .  1  .  .  .
 91  .  .  1  .  .  .  1  .  .  .  .  .  .  1  .  .  .  .  3  1  .  1  .  .  .  .  .  1  .  .
 93  .  .  .  .  .  .  .  .  .  1  1  1  .  .  1  .  .  1  .  2  .  .  1  .  .  .  .  .  1  .
103  .  1  .  .  1  .  1  .  .  .  .  .  .  .  .  .  .  .  .  .  4  .  .  .  2  .  .  .  .  .
107  .  .  1  1  .  .  .  .  .  .  .  .  .  .  .  .  .  .  .  .  2  3  1  .  .  .  1  .  .  .
109  .  .  .  1  1  .  .  .  .  .  .  .  1  .  .  .  .  .  1  .  .  .  3  1  .  .  .  1  .  .
117  .  .  .  .  .  1  .  .  .  .  1  .  .  1  .  .  1  .  .  1  1  .  .  2  .  .  1  .  .  .
127  .  .  .  .  .  .  .  2  .  .  .  .  .  .  .  .  .  .  .  .  .  .  .  .  6  .  .  .  .  1
171  .  .  .  .  .  1  .  .  .  .  .  .  .  .  .  .  .  1  1  .  .  2  1  1  .  2  .  .  .  .
191  .  .  .  .  .  .  .  .  .  .  .  1  .  .  1  .  .  .  .  .  .  .  .  .  2  .  4  1  .  .
223  .  .  .  .  .  .  .  1  .  .  .  .  1  1  .  .  .  .  .  .  .  .  .  .  .  .  1  4  1  .
239  .  .  .  .  .  .  .  .  .  .  .  1  1  1  1  .  .  .  .  .  .  .  .  .  .  .  .  1  4  .
511  .  .  .  .  .  .  .  .  .  .  .  .  .  .  .  .  .  .  .  .  .  .  .  .  .  .  .  .  .  9
    -------------------------------------------------------------------------------------------          
\end{verbatim}
\normalsize
\label{TMRuleHA}
\end{table}   

\newpage
\subsection{Rule HB}

Rule \textit{HB} is neighborhood-symmetric. 
The transition matrices are given in Table 
\ref{TMRuleHB}. 
The indexes of the matrices (I, II, IV) represent state equivalents under shift,
and the indexes of matrices (I, III, V)              
represent state equivalents under shift and mirror. 
The $(i,j)$ entries are the probabilities $p(i,j)$ to be divided by $n$ to change from state $i$ to $j$.

\begin{table}[H]
\center
\caption{
Rule \textit{HB} transition matrices for $n=5, 7, 9$. 
(II) The state 13 (the mirror of 11) is included in the matrix, and
(III) the state 13 is excluded.
(IV) The 29 indexes of the matrix are state representatives of equivalents under shift. 
(V) The 22 indexes are representatives of state equivalents under shift  and mirror.
}
\tiny
\begin{verbatim}
(I) (n = 5)                    (II) (n = 7)                                      (III) (n = 7) 
             1  7  11 31                     1  7  11 13 19 21 31 47 55 127            1  7  11 19 21 31 47 55 12      
            ------------                    -------------------------------           ---------------------------
00001 =  1   3  2  0  0        0000001 =  1  5  2  .  .  .  .  .  .  .  .           1  5  2  .  .  .  .  .  .  .
00111 =  7   1  2  0  2        0000111 =  7  2  3  .  .  .  .  2  .  .  .           7  2  3  .  .  .  2  .  .  .
01011 = 11   1  0  4  0        0001011 = 11  1  .  4  .  .  .  .  1  1  .          11  1  .  4  .  .  .  1  1  .
11111 = 31   0  0  0  5        0001101 = 13  1  .  .  4  .  .  .  1  1  .          19  1  .  .  4  .  2  .  .  .
            ------------       0010011 = 19  1  .  .  .  4  .  2  .  .  .          21  .  .  .  .  6  .  1  .  .
                               0010101 = 21  .  .  .  .  .  6  .  1  .  .          31  .  2  .  .  .  4  .  .  1
                               0011111 = 31  .  2  .  .  .  .  4  .  .  1          47  .  .  2  .  .  .  5  .  .
                               0101111 = 47  .  .  1  1  .  .  .  5  .  .          55  .  1  2  .  .  .  .  4  .
                               0110111 = 55  .  1  1  1  .  .  .  .  4  .         127  .  .  .  .  .  .  .  .  7
                               1111111 =127  .  .  .  .  .  .  .  .  .  7             ---------------------------
                                            -------------------------------             
                         
(IV) (n = 9)                                                        107   117   171   223   511
     1  7  11 13 19 21 25 31 35 37 41 47 55 59 61 73 79 87 91 93 103   109   127   191   239
    -------------------------------------------------------------------------------------------  
  1  7  2  .  .  .  .  .  .  .  .  .  .  .  .  .  .  .  .  .  .  .  .  .  .  .  .  .  .  .  .
  7  2  5  .  .  .  .  .  2  .  .  .  .  .  .  .  .  .  .  .  .  .  .  .  .  .  .  .  .  .  .
 11  1  .  6  .  .  .  .  .  .  .  .  1  .  1  .  .  .  .  .  .  .  .  .  .  .  .  .  .  .  .
 13  1  .  .  6  .  .  .  .  .  .  .  .  1  .  1  .  .  .  .  .  .  .  .  .  .  .  .  .  .  .
 19  1  .  .  .  5  .  .  1  .  .  .  .  .  .  .  .  1  .  .  .  1  .  .  .  .  .  .  .  .  .
 21  .  .  .  .  .  7  .  .  .  .  .  .  .  .  .  .  .  1  .  .  .  .  .  1  .  .  .  .  .  .
 25  1  .  .  .  .  .  5  1  .  .  .  .  .  .  .  .  1  .  .  .  1  .  .  .  .  .  .  .  .  .
 31  .  2  .  .  .  .  .  5  .  .  .  .  .  .  .  .  .  .  .  .  .  .  .  .  2  .  .  .  .  .
 35  1  .  .  .  .  .  .  .  4  .  .  1  1  1  1  .  .  .  .  .  .  .  .  .  .  .  .  .  .  .
 37  .  .  .  .  .  .  .  .  .  6  .  .  .  .  1  .  .  1  .  1  .  .  .  .  .  .  .  .  .  .
 41  .  .  .  .  .  .  .  .  .  .  6  1  .  .  .  .  .  .  .  1  .  .  .  1  .  .  .  .  .  .
 47  .  .  1  .  .  .  .  .  1  .  .  5  .  .  .  .  .  .  .  .  .  .  .  .  .  .  1  .  1  .
 55  .  1  .  1  .  .  .  .  1  .  .  .  4  .  .  .  .  .  .  .  .  .  .  .  .  .  .  1  1  .
 59  .  1  1  .  .  .  .  .  1  .  .  .  .  4  .  .  .  .  .  .  .  .  .  .  .  .  .  1  1  .
 61  .  .  .  1  .  .  .  .  1  .  .  .  .  .  5  .  .  .  .  .  .  .  .  .  .  .  1  .  1  .
 73  .  .  .  .  .  .  .  .  .  .  .  .  .  .  .  6  3  .  .  .  .  .  .  .  .  .  .  .  .  .
 79  .  .  .  .  1  .  1  .  .  .  .  .  .  .  .  .  5  .  .  .  .  .  .  .  2  .  .  .  .  .
 87  .  .  .  .  .  1  .  .  .  1  .  .  .  .  .  .  .  6  .  .  .  .  .  .  .  .  1  .  .  .
 91  .  .  1  .  .  .  1  .  .  .  .  .  .  .  .  .  .  .  6  .  .  .  .  .  .  .  .  1  .  .
 93  .  .  .  .  .  .  .  .  .  1  1  .  .  .  .  .  .  .  .  6  .  .  .  .  .  .  .  .  1  .
103  .  1  .  .  1  .  1  .  .  .  .  .  .  .  .  .  .  .  .  .  4  .  .  .  2  .  .  .  .  .
107  .  .  1  1  .  .  .  .  .  .  .  .  .  .  .  .  .  .  .  .  .  6  .  .  .  .  1  .  .  .
109  .  .  .  1  1  .  .  .  .  .  .  .  .  .  .  .  .  .  .  .  .  .  6  .  .  .  .  1  .  .
107  .  .  .  .  .  1  .  .  .  .  1  .  .  .  .  .  .  .  .  .  .  .  .  6  .  .  1  .  .  .
127  .  .  .  .  .  .  .  2  .  .  .  .  .  .  .  .  .  .  .  .  .  .  .  .  6  .  .  .  .  1
171  .  .  .  .  .  1  .  .  .  .  .  .  .  .  .  .  .  .  .  .  .  .  .  .  .  8  .  .  .  .
191  .  .  .  .  .  .  .  .  .  .  .  1  .  .  1  .  .  .  .  .  .  .  .  .  .  .  7  .  .  .
223  .  .  .  .  .  .  .  1  .  .  .  .  1  1  .  .  .  .  .  .  .  .  .  .  .  .  .  6  .  .
239  .  .  .  .  .  .  .  .  .  .  .  1  1  1  1  .  .  .  .  .  .  .  .  .  .  .  .  .  5  .
511  .  .  .  .  .  .  .  .  .  .  .  .  .  .  .  .  .  .  .  .  .  .  .  .  .  .  .  .  .  9            
    -------------------------------------------------------------------------------------------  
                          
(V) (n = 9)                                          107   171   223   511
     1  7  11 19 21 31 35 37 47 55 73 79 87 91 93 103   127   191   239
    ----------------------------------------------------------------------
  1  7  2  .  .  .  .  .  .  .  .  .  .  .  .  .  .  .  .  .  .  .  .  .
  7  2  5  .  .  .  2  .  .  .  .  .  .  .  .  .  .  .  .  .  .  .  .  .
 11  1  1  4  1  .  .  .  .  1  1  .  .  .  .  .  .  .  .  .  .  .  .  .
 19  1  .  1  3  .  1  1  .  .  .  .  1  .  .  .  1  .  .  .  .  .  .  .
 21  .  .  2  2  1  .  .  2  .  .  .  .  2  .  .  .  .  .  .  .  .  .  .
 31  .  2  .  .  .  5  .  .  .  .  .  .  .  .  .  .  .  2  .  .  .  .  .
 35  1  .  .  2  .  .  2  .  2  2  .  .  .  .  .  .  .  .  .  .  .  .  .
 37  .  .  .  1  1  .  1  2  1  .  1  .  1  .  1  .  .  .  .  .  .  .  .
 47  .  .  1  .  .  1  1  .  2  1  .  1  .  .  .  .  .  .  .  1  .  1  .
 55  .  1  1  .  .  .  1  .  1  3  .  .  .  .  .  .  .  .  .  .  1  1  .
 73  .  .  .  .  .  .  .  6  .  .  .  3  .  .  .  .  .  .  .  .  .  .  .
 79  .  .  .  2  .  .  .  .  2  .  .  3  .  .  .  .  .  2  .  .  .  .  .
 87  .  .  .  .  1  .  .  1  .  1  .  1  1  1  1  1  .  .  .  1  .  .  .
 91  .  .  1  1  .  .  .  .  .  1  .  .  1  2  1  .  1  .  .  .  1  .  .
 93  .  .  .  .  .  .  .  2  2  .  .  .  2  2  .  .  .  .  .  .  .  1  .
103  .  1  .  2  .  .  .  .  .  .  .  .  .  .  .  4  .  2  .  .  .  .  .
107  .  .  2  .  .  .  .  .  .  .  .  .  .  2  .  2  2  .  .  1  .  .  .
127  .  .  .  .  .  2  .  .  .  .  .  .  .  .  .  .  .  6  .  .  .  .  1
171  .  .  .  .  1  .  .  .  .  .  .  .  2  2  .  .  2  .  2  .  .  .  .
191  .  .  .  .  .  .  .  .  2  .  .  .  .  .  .  .  .  2  .  3  2  .  .
223  .  .  .  .  .  1  .  .  .  2  .  .  .  .  .  .  .  .  .  2  2  2  .
239  .  .  .  .  .  .  .  .  2  2  .  .  .  .  .  .  .  .  .  .  2  3  .
511  .  .  .  .  .  .  .  .  .  .  .  .  .  .  .  .  .  .  .  .  .  .  9 
    ---------------------------------------------------------------------                            
\end{verbatim}
\normalsize
\label{TMRuleHB} 
\end{table}   

\newpage
\subsection{Rule HC}
Rule \textit{HC} is not neighborhood-symmetric. The transition matrices are given in Table 
\ref{TMRuleHC}. The indexes
of the matrices represent state equivalents under shift. The $(i, j)$ entries are the probabilities
$p(i, j)$ to be divided by $n$ to change from state $i$ to $j$.

\begin{table}[H]
\center
\caption{
Rule \textit{HC} transition matrices for $n=5, 7, 9$. 
 The indexes of the matrix are state representatives of equivalents under shift. 
}
\tiny
\begin{verbatim}
(I) (n = 5)                      (II) (n = 7)
             1  7  11 31                        1  7  11 13 19 21 31 47 55 127      
            ------------                       _______________________________  
00001 =  1   4  1  0  0          0000001 =  1   6  1  .  .  .  .  .  .  .  .    
00111 =  7   1  3  0  1          0000111 =  7   1  5  .  .  .  .  1  .  .  .    
01011 = 11   1  0  4  0          0001011 = 11   1  .  5  .  .  .  .  1  .  .    
11111 = 31   0  0  0  5          0001101 = 13   1  .  .  5  .  .  .  .  1  .    
            ------------         0010011 = 19   1  .  .  .  4  .  2  .  .  .    
                                 0010101 = 21   .  .  .  .  .  6  .  1  .  .    
                                 0011111 = 31   .  1  .  .  .  .  5  .  .  1    
                                 0101111 = 47   .  .  .  1  .  .  .  6  .  .    
                                 0110111 = 55   .  1  1  .  .  .  .  .  5  .    
                                 1111111 =127   .  .  .  .  .  .  .  .  .  7    
                                               ------------------------------

(III) (n = 9)                                                     107   117   171   223   511 
      7  11 13 19 21 25 31 35 37 41 47 55 59 61 73 79 87 91 93 103   109   127   191   239
     -------------------------------------------------------------------------------------------
   1  8  1  .  .  .  .  .  .  .  .  .  .  .  .  .  .  .  .  .  .  .  .  .  .  .  .  .  .  .  .
   7  1  7  .  .  .  .  .  1  .  .  .  .  .  .  .  .  .  .  .  .  .  .  .  .  .  .  .  .  .  .
  11  1  .  7  .  .  .  .  .  .  .  .  1  .  .  .  .  .  .  .  .  .  .  .  .  .  .  .  .  .  .
  13  1  .  .  7  .  .  .  .  .  .  .  .  1  .  .  .  .  .  .  .  .  .  .  .  .  .  .  .  .  .
  19  1  .  .  .  6  .  .  1  .  .  .  .  .  .  .  .  1  .  .  .  .  .  .  .  .  .  .  .  .  .
  21  .  .  .  .  .  8  .  .  .  .  .  .  .  .  .  .  .  1  .  .  .  .  .  .  .  .  .  .  .  .
  25  1  .  .  .  .  .  6  1  .  .  .  .  .  .  .  .  .  .  .  .  1  .  .  .  .  .  .  .  .  .
  31  .  1  .  .  .  .  .  7  .  .  .  .  .  .  .  .  .  .  .  .  .  .  .  .  1  .  .  .  .  .
  35  1  .  .  .  .  .  .  .  6  .  .  .  .  1  1  .  .  .  .  .  .  .  .  .  .  .  .  .  .  .
  37  .  .  .  .  .  .  .  .  .  7  .  .  .  .  1  .  .  .  .  1  .  .  .  .  .  .  .  .  .  .
  41  .  .  .  .  .  .  .  .  .  .  7  1  .  .  .  .  .  .  .  .  .  .  .  1  .  .  .  .  .  .
  47  .  .  .  .  .  .  .  .  1  .  .  7  .  .  .  .  .  .  .  .  .  .  .  .  .  .  1  .  .  .
  55  .  1  .  .  .  .  .  .  1  .  .  .  6  .  .  .  .  .  .  .  .  .  .  .  .  .  .  1  .  .
  59  .  1  1  .  .  .  .  .  .  .  .  .  .  6  .  .  .  .  .  .  .  .  .  .  .  .  .  .  1  .
  61  .  .  .  1  .  .  .  .  .  .  .  .  .  .  7  .  .  .  .  .  .  .  .  .  .  .  .  .  1  .
  73  .  .  .  .  .  .  .  .  .  .  .  .  .  .  .  6  3  .  .  .  .  .  .  .  .  .  .  .  .  .
  79  .  .  .  .  .  .  1  .  .  .  .  .  .  .  .  .  6  .  .  .  .  .  .  .  2  .  .  .  .  .
  87  .  .  .  .  .  .  .  .  .  1  .  .  .  .  .  .  .  7  .  .  .  .  .  .  .  .  1  .  .  .
  91  .  .  1  .  .  .  1  .  .  .  .  .  .  .  .  .  .  .  6  .  .  .  .  .  .  .  .  1  .  .
  93  .  .  .  .  .  .  .  .  .  .  1  .  .  .  .  .  .  .  .  7  .  .  .  .  .  .  .  .  1  .
 103  .  1  .  .  1  .  .  .  .  .  .  .  .  .  .  .  .  .  .  .  5  .  .  .  2  .  .  .  .  .
 107  .  .  1  1  .  .  .  .  .  .  .  .  .  .  .  .  .  .  .  .  .  6  .  .  .  .  1  .  .  .
 109  .  .  .  1  1  .  .  .  .  .  .  .  .  .  .  .  .  .  .  .  .  .  6  .  .  .  .  1  .  .
 117  .  .  .  .  .  1  .  .  .  .  .  .  .  .  .  .  .  .  .  .  .  .  .  7  .  .  1  .  .  .
 127  .  .  .  .  .  .  .  1  .  .  .  .  .  .  .  .  .  .  .  .  .  .  .  .  7  .  .  .  .  1
 171  .  .  .  .  .  1  .  .  .  .  .  .  .  .  .  .  .  .  .  .  .  .  .  .  .  8  .  .  .  .
 191  .  .  .  .  .  .  .  .  .  .  .  .  .  .  1  .  .  .  .  .  .  .  .  .  .  .  8  .  .  .
 223  .  .  .  .  .  .  .  1  .  .  .  .  .  1  .  .  .  .  .  .  .  .  .  .  .  .  .  7  .  .
 239  .  .  .  .  .  .  .  .  .  .  .  1  1  .  .  .  .  .  .  .  .  .  .  .  .  .  .  .  7  .
 511  .  .  .  .  .  .  .  .  .  .  .  .  .  .  .  .  .  .  .  .  .  .  .  .  .  .  .  .  .  9             
    -------------------------------------------------------------------------------------------              
\end{verbatim}
\normalsize
\label{TMRuleHC}
\end{table}   


\begin{thebibliography}{99.}
\bibitem{2024Fates}
Nazim Fatès. 
\textit{Asynchronous Cellular Systems that Solve the Parity Problem.}
International Workshop on Cellular Automata and Discrete Complex Systems. 
Cham: Springer Nature Switzerland, 2024.

\bibitem{2019Ruivo}
Enrico L.P. Ruivo and Pedro P.B. de Oliveira. 
\textit{A perfect solution to the parity problem with elementary cellular automaton 150 under asynchronous update.} 
Information Sciences, 493:138-–151, 2019. doi:10.1016/j.ins.2019.04.045.

\bibitem{2025Wolnik}
Barbara Wolnik, Anna Nenca, Pedro Paulo Balbi, and Bernard De Baets.
\textit{Cellular automata can really solve the parity problem.}
arXiv preprint arXiv:2501.08684 (2025)

\bibitem{2013Betel}
Heather Betel, Pedro P. B. de Oliveira, and Paola Flocchini. 
\textit{Solving the parity problem in one-dimensional cellular automata.} 
Natural Computing, 12(3):323–337, 2013. doi:10.1007/
s11047-013-9374-9.

\bibitem{1998Sipper}
Moshe Sipper. Computing with cellular automata: Three cases for nonuniformity. Physical
Review E, 57:3589–3592, 1998. doi:10.1103/PhysRevE.57.3589.

\end{thebibliography}
\end{document}